\documentclass[11pt, twocolumn]{article}

\usepackage[english]{babel}
\usepackage[dvipsnames]{xcolor}
\usepackage{multirow}
\usepackage{caption}
\usepackage[letterpaper, top=0.5in, bottom=0.5in, left=0.5in, right=0.5in, marginparwidth=1in]{geometry}

\usepackage{amsmath}
\usepackage{graphicx}
\usepackage[colorlinks=true, allcolors=blue]{hyperref}
\usepackage{titlesec}
\titleformat{\section}
{\color{Red}\normalfont\large\bfseries\sffamily}
{\color{Red}\thesection}{1em}{}

\titleformat{\subsection}
{\color{Red}\normalfont\normalsize\bfseries\sffamily}
{\color{Red}\thesubsection}{1em}{}

\titleformat{\subsubsection}
{\color{Red}\normalfont\small\bfseries\sffamily}
{\color{Red}\thesubsubsection}{1em}{}

\title{\textbf{Modeling non-genetic information dynamics in cells using reservoir computing}}
\date{\vspace{-5ex}}
\author{Dipesh Niraula$^1$\footnote{Corresponding Author: Dipesh.Niraula@moffitt.org}, Issam El Naqa$^1$, Jack Adam Tuszynski$^{2,3,4}$, \& Robert A. Gatenby$^5$\\
{\small $^1$ Department of Machine Learning, Moffitt Cancer Center, Tampa, FL, USA},\\
{\small $^2$ Departments of Physics and Oncology, University of Alberta, Edmonton, AB, CAN},\\
{\small $^3$ Department of Data Science and Engineering, The Silesian University of Technology, Gliwice, 44-100, Poland},\\{\small $^4$ Department of Mechanical and Aerospace Engineering, Politecnico di Torino, Turin, Italy, I-10129},\\
{\small $^5$ Departments of Radiology and Integrated Mathematical Oncology, Moffitt Cancer Center, Tampa, FL, USA}}

\begin{document}
\maketitle
\begin{abstract}

Virtually all cells use energy and ion-specific membrane pumps to maintain large transmembrane gradients of Na$^{+}$, K$^{+}$, Cl$^{-}$, Mg$^{++}$, and Ca$^{++}$.  Although they consume up to 1/3 of a cell’s energy budget, the corresponding evolutionary benefit of transmembrane ion gradients remain unclear. Here, we propose that ion gradients enable a dynamic and versatile biological system that acquires, analyzes, and responds to environmental information. We hypothesize environmental signals are transmitted into the cell by ion fluxes along pre-existing gradients through gated ion-specific membrane channels.  The consequent changes of cytoplasmic ion concentration can generate a local response and orchestrate global or regional responses through wire-like ion fluxes along pre-existing and self-assembling cytoskeleton to engage the endoplasmic reticulum, mitochondria, and nucleus.  

Here, we frame our hypothesis through a quasi-physical (Cell-Reservoir) model that treats intra-cellular ion-based information dynamics as a sub-cellular process permitting spatiotemporally resolved cellular response that is also capable of learning complex nonlinear dynamical cellular behavior. We demonstrate the proposed ion dynamics permits rapid dissemination of response to information extrinsic perturbations that is consistent with experimental observations. 
\end{abstract}


\section{Introduction}
Investigation of information dynamics necessary for living systems is often limited to the genomic encoding processes \cite{1}. However, information in the genome is fixed and requires time to transcribe while optimal evolutionary fitness \cite{2,3} demands cells to continuously adapt to diverse, changing environmental conditions including rapid response to sudden life-threatening perturbations.  We hypothesize flexible and rapid response to changes in the environment require rapid communication to and from the plasma membrane, which is the cell’s primary interface with the external environment. Thus, while the genome provides the Darwinian mechanism of inheritance, we hypothesize \cite{4} there is a secondary information system, built by the inherited macromolecular backbone, that is necessary to allow cells to adapt to changing environmental conditions – a key component of Darwinian fitness.  

Here, we theoretically investigate this proposed secondary information system that is likely is built upon the transmembrane ion gradient \cite{5}. These gradients globally result in a transmembrane potential that is well recognized as a critical factor in cellular proliferation and differentiation, although the mechanisms of this process remain unknown \cite{6, 7}. Furthermore, recent studies have demonstrated that the size of the transmembrane potential exhibits extensive spatial and temporal variations across the cell surface \cite{7,8, 9} that must represent local transmembrane ion fluxes through gated channels. 

\begin{figure*}
    \centering
    \includegraphics[width=0.9\textwidth]{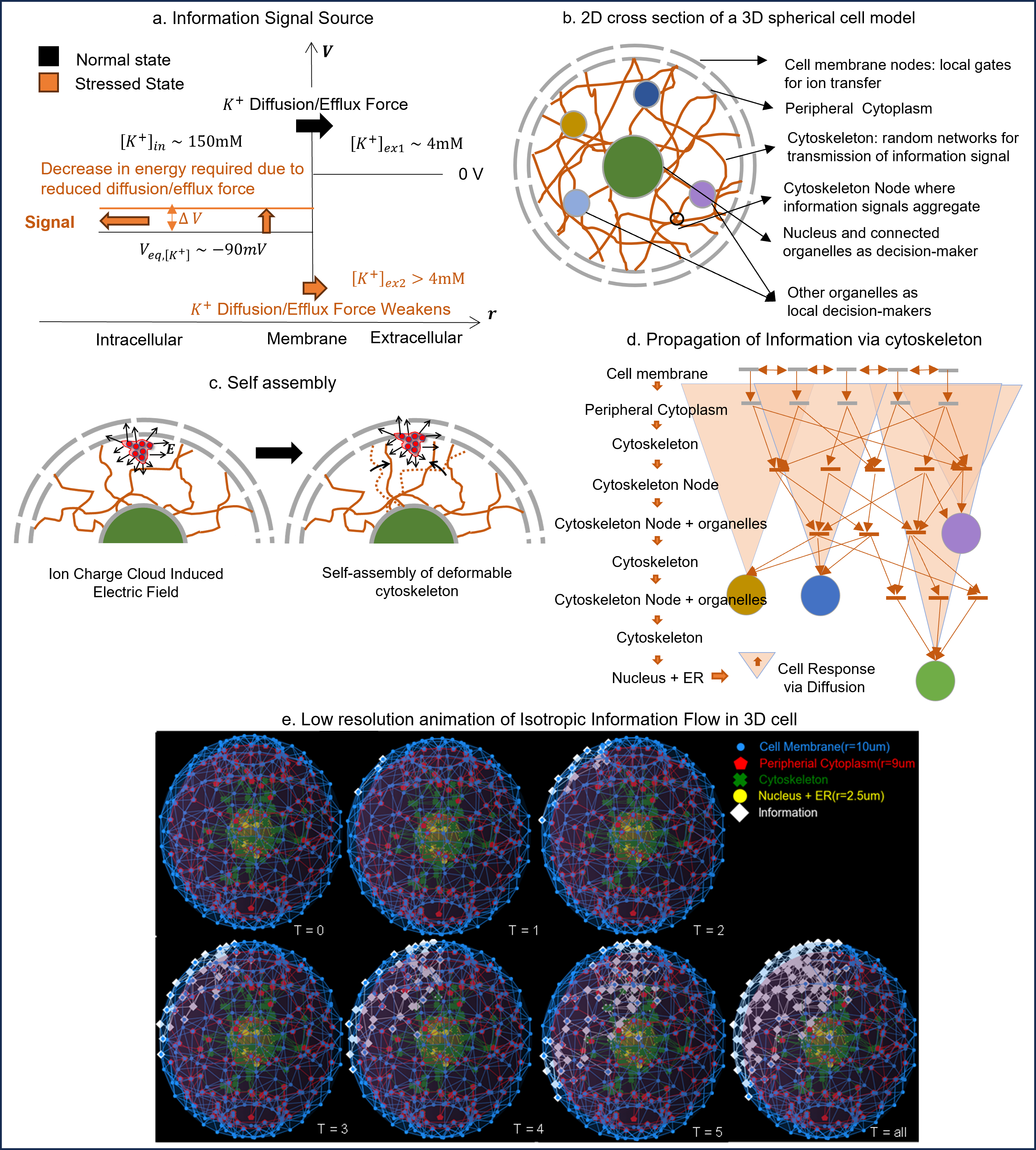}
    \caption{\small \textbf{Intracellular information dynamics model}. \textbf{a.} Change in external potassium ion concentration changes potassium diffusive/efflux force, which changes the energy required to maintain the internal potassium concentration resulting in an information signal that gets passed into the cell via conducting cytoskeleton \textbf{b.} Two-dimensional cross-section of a three-dimensional spherical cell model consisting of cell membrane, peripheral cytoplasm, cell-organelles, and cytoskeleton for minimally capturing the intracellular information dynamics. \textbf{c.} Change in local ion-concentration can result in re-assembly of cytoskeletons.  \textbf{d.} Mechanism for propagation of information via the cytoskeleton for inducing appropriate cell response. \textbf{e.} Low-resolution animation of information dynamics where the information took 5-time steps (each time step is simulated to be in the range of 2-20$\mu$s) to travel from cell boundary to the central organelle through a random network of cytoskeletons.}
    \label{fig:fig1}
\end{figure*}

Here, we focus explicitly on the role of transmembrane ion fluxes as carriers of information. These dynamics are well recognized in traveling depolarization waves of neurons described by Hodgkin and Huxley \cite{10}.  We hypothesize ion flows in neurons represent a specialized adaptation of a broader but generally unrecognized information acquisition, processing, and response dynamics centered around the cell membrane. By coupling specific receptor proteins as gates in ion channels, environmental information received by the receptor causes the gate to open thus transmitting the information into the cell by ion fluxes (along the pre-existing transmembrane gradient) resulting in local changes in cytoplasmic ion concentrations. This information, in the form of local changes in cytoplasmic ion concentrations, is “processed” through ion-specific changes in protein functions. For example, there are about 300 magnesium-activated enzymes, many of which are related to glucose metabolism \cite{11}, pH response, and cell adhesion \cite{12}. Similarly, K$^+$ concentrations regulate apoptotic enzymes \cite{13, 14} and aldehyde dehydrogenases \cite{15}, and Ca$^{++}$ concentrations govern activity of some matrix metalloproteinases \cite{16} and cellular responses to extrinsic ligands.  Furthermore, local efflux of  K$^+$, because it is the dominant mobile cation in cytoplasm, will reduce shield of negative charges on the inner leaf of the cell membrane permitting a small and transient electric field that may  attract nearby positively charged regions of macromolecules \cite{4}.

Thus, ion-based information processing at the membrane permits rapid and local responses to focal and transient signal or perturbations from the environment.  However, optimal cell fitness also requires a regional or global cellular response to high frequency or high amplitude perturbations. How is this information transmitted to the other components of the cell?  One mechanism, like Hodgkin Huxley dynamics, is through a propagating ion wave to adjacent regions of the membrane. Thus, propagating calcium waves are seen in, for example, following mechanical stimulation of keratinocytes \cite{17}.

In addition, we propose large perturbations or sustained changes in the ion cytoplasmic concentration adjacent to the membrane can be transmitted to other cellular organelles via self-organizing elements of the cytoskeleton. Changes in ion concentration adjacent to the membrane promote cytoskeleton self-assembly. Once formed, the difference in ion concentration at the peripheral end of the microfilament or microtubule, compared to its proximal end, results in flow of ions or electrons along the filament. This is consistent with an extensive experimental literature demonstrating ion and electron conductance within all components of the cytoskeleton \cite{18,19,20,21,22,23,24,25}. Interestingly, while this conductance has been extensively investigated experimentally, little if any work has addressed its biological function. In our model, the cytoskeleton both transmits signals and facilitates the cellular response to external information. Thus, for example, cytoskeleton is deeply enmeshed and communicates with mitochondria and endoplasmic reticulum \cite{26} to deliver energy and macromolecules to the site of perturbation. Furthermore, the cytoskeleton is linked to the nuclear membrane through KASH (Klarsicht, ANC-1, Syne homology) and SUN  (Sad1 and UNC-84) proteins (collectively described as the LINC complex (linker of nucleoskeleton and cytoskeleton), which can control gene transcription and chromosomal movement \cite{27,28} thus directly linking the genome with other cellular components \cite{29}.
 
Our hypothesis is based upon the following observations: 1. The well documented ability of microtubules and microfilaments to conduct ions and electron \cite{18,19,21,23,25,30,31,32,33,34,35,36}. 2. The dynamics of cytoskeleton self-assembly \cite{37,38,39,40,41} and the observed dependence of the effects of cytoplasmic ion concentrations and \cite{42} membrane potential on these dynamics. 3. Evidence of strong mutual interactions of elements of the cytoskeleton and ion channels \cite{43,44,45,46,47,48,49,50,51}. 4. The deep interconnection of the cytoskeleton \cite{43,44,49,52,53,54,55} with mitochondria and endoplasmic reticulum \cite{26} to deliver energy and macromolecules to the site of perturbation and the nuclear membrane through LINC complex, which can control gene transcription and chromosomal movement \cite{27,28}. 

In prior modeling studies, we have investigated the information dynamics of transmembrane flux \cite{5} and ion flow along cytoskeleton \cite{56}. Here, we frame our global hypothesis of cellular information acquisition, processing, and decision making through a quasi-physical model of cell (Cell-Reservoir) that integrates information flow from membrane ion fluxes through existing and self-assembling elements of the cytoskeleton. We demonstrate this system permits rapid, spatiotemporally resolved transmission of environmental information to critical response elements in the cell that are capable of learning. We demonstrate this proposed network allows both rapid dissemination of and response to information extrinsic perturbations consistent with experimental observations. 

Living cells are non-linear dynamic systems that must constantly adapt to their environment. Modeling intracellular information dynamics solely based on physical and chemical laws needs many approximations and can still fall short in capturing the complexity of subcellular processes. Thus, we developed a quasi-physical reservoir computing framework that pairs a graph-based reservoir system with decision-making model for modeling cellular decision-making processes capable of learning directly from measurements. The novelty of Cell-Reservoir is that it consists of randomly generated networks of conducting cytoskeletons capable of transmitting electrical signals from the cell membrane to the various organelles, which are considered in this model as the local centers of decision-making. The signal transmission follows Ohms and Kirchhoff’s current laws in an intracellular ionic potential landscape dictated by the screened Poisson equation with Debye-Huckel approximation \cite{57}. The flow of information follows causality principles similar to Huygen’s principle where each point acts as source of information for later time step naturally giving rise to spatiotemporally resolved properties. Computationally, the Cell-Reservoir assumes voltage and current as node properties and conductivity as edge properties, which is stored in spatially correct location in a grid-graph data structure, reducing both space and time complexity. In addition, each node stores a physical signal (current) and a memory signal (exponential time averages of physical signal), enabling our model to effectively have short-term memory of environmental perturbations and to learn dynamical cell behavior. Finally, owing to the complexity of cellular decision-making processes that involve an ensemble of subcellular processes and events, we employ a data-driven decision-making model which can learn from available cellular measurements. The advantage of our approach is its flexibility, which allows us to add new layers of decision-making processes depending on the availability of such measurements. We demonstrate that Cell-Reservoir’s ability and flexibility, via supervised learning of CD8 T-cell’s behavior activated by different concentration of CD3 and CD28 T-cell receptor protein in the presence of different levels of extracellular K$^+$ concentration. Additionally, we show the directed nature of information dynamics resulting from the spatiotemporal properties of Cell-Reservoir, present results from perturbation study where we investigated the minimum cytoskeleton volume required to guarantee signal transmission from cell membrane to central organelle along with a visual inspection via uniform manifold approximation and projection (UMAP) clustering of the cell geometry and carried out noise analysis by investigating Cell Reservoirs’ learning ability for varying signal-to-noise ratios (SNRs).
\begin{table*}
\footnotesize
\caption{\small Typical cellular ion concentrations and corresponding Nernst equilibrium potential}\label{tab:eqpot} 
\centering
\begin{tabular}{ccccc}
\hline
\multirow{2}{*}{Ion}&\multirow{2}{*}{Intracellular concentration (mM)}&\multirow{2}{*}{Extracellular concentration (mM)}&\multicolumn{2}{c}{Equilibrium Potential (mV)}\\
	&		&		&	$T=21^{\circ}C$	&	$T=37^{\circ}C$	\\
\hline
Na$^+$	&	13	&	142	&	60.6	&	63.9	\\
Cl$^-$	&	5	&	120	&	-80.6	&	-84.9	\\
K$^+$	&	150	&	4	&	-91.9	&	-96.9	\\
Ca$^{++}$	&	0.0001	&	1	&	116.7	&	123.1	\\
Mg$^{++}$	&	1	&	0.5	&	-8.8	&	-9.3	\\
HCO3$^-$	&	8	&	27	&	-30.8	&	-32.5	\\
Organic anions	&	155	&	0* &	-	&	-	\\
\hline
\multicolumn{5}{l}{\footnotesize *will be non-zero in the extracellular space of multicellular organisms}
\end{tabular}
\end{table*}

\begin{figure*}
    \centering
    \includegraphics[width=0.9\textwidth]{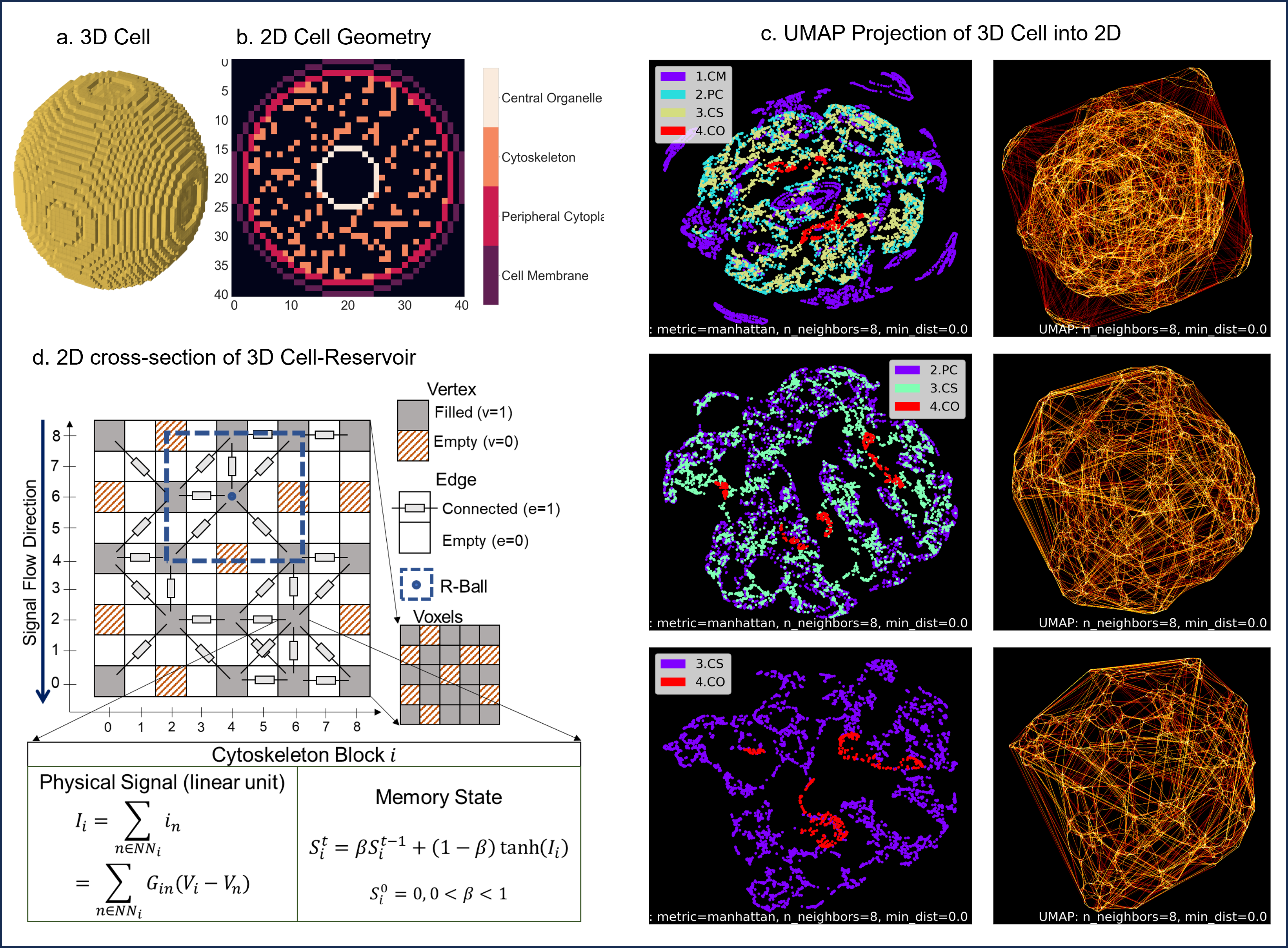}
    \caption{\small \textbf{Cell Geometry and Cell-Reservoir}. \textbf{a.} 3D spherical cell composed of voxels. \textbf{b.} 2D cross section of cell showing cell membrane (CM), peripheral cytoplasm (PC), cytoskeleton (CS), and central organelle (CO). \textbf{c.} UMAP projection of 3D cell into 2D Space, where the three rows correspond to projections on same cell with all components, without CM, and without CM and PC, respectively, while the right column shows the UMAP nearest neighbor (NN) connection. \textbf{d.} Cell-Reservoir $G(V,E)$ discretizes physical 3D space into $n\times n\times n$ voxels (vertices) by assigning them to the even indices $(2i,2j,2k)$ of a 3D tensor of size $2n-1 \times 2n-1 \times 2n-1$.  Each voxel is surrounded by 26 edges. A voxel is either empty ($v=0$) or filled ($v=1$). A cubic R-ball of size $5\times 5\times 5$ can traverse through Cell-Reservoir for locating the NN and their inter-relationships. Each filled vertex has two states that stores the physical signal $I_i$ and the memory signal $S_i$.}
    \label{fig:fig2}
\end{figure*}
\section{Methods}
\subsection{The Biology of Intracellular Information Flow}
Cells are electrochemically active systems that constantly spend energy to maintain a fixed intracellular ion concentration (Table \ref{tab:eqpot}). This non-random distribution of ions across the cell membrane has been estimated to contain about 10$^11$ bits of Shannon information in E coli and substantially more in large eukaryotic cells, which also maintain gradients across the membranes of intracellular organelles. This action potential gradient information is used in processing environmental information as it impacts gated ion-specific channels in the membrane. When external perturbations open the gate, information in the form of rapid ion fluxes flows into or out of the cell. Local changes in the ion concentration of the cytoplasm alter function and localization of intra-membrane and peripheral membrane proteins allowing analysis and response to perturbation.

The cellular cytoplasm contains a network of interconnected cytoskeleton (microtubules and microfilaments) that self-assemble in response to variations in local cytoplasmic ion concentrations and external forces. In our model, the cytoskeleton integrates biomechanical function with information transmission via flow of ion and physical stresses. That is, information through transmembrane ion fluctuations change concentration in the local cytoplasm and can prompt local cytoskeleton self-assembly.  These microfilaments and microtubules can transmit information via wire-like flow of ions and, through their biomechanical function, mediate a response to local perturbations.  Mitochondria and endoplasmic reticulum, closely linked to each other via cytoskeleton, can be signaled to increase production of energy and macromolecules and moved closer to the site of perturbation by the cytoskeleton. Similarly, the cytoskeleton connects with the nuclear membrane via the LINK complex which affects gene expression and chromosome localization. 

In summary, an external perturbation induces opening of gated membrane channels allowing specific ions to flow along concentration gradients into or out of cell, decreasing negativity of the local cellular electrochemical potential and altering ion concentrations in the adjacent cytoplasm. Once the perturbation resolves, the membrane channel will close initiating repolarization via the membrane pumps. 

Here, the local cellular response (i.e., at the site of the perturbation) is determined by altered localization and the functions of intramembrane and peripheral membrane proteins due to local ion fluxes.  For larger and prolonged perturbations, changes in ion concentrations in the peripheral cytoplasm are greater in amplitude and diffuse farther into the cytoplasm. These induce local cytoskeleton self-assembly, which can then conduct ions to intracellular organelles (mitochondria, endoplasmic reticulum, and nucleus) to generate a more global cellular response with increased production of energy and macromolecules from the mitochondria and endoplasmic reticulum/Golgi complex with transport of organelles to the site of maximal perturbation.   Ion signals can flow along the cytoskeleton to the LINK complex on the nuclear membrane, which can initiate changes in gene transcription and chromosomal localization.

\subsection{The Physics of Intracellular Information Dynamics}
Healthy cells maintain a certain intracellular potassium ion concentration ($[K^+]_{in}$) which is much higher than the extracellular potassium ion concentration ($[K^+]_{ex}$). To maintain such concentration gradient the cell must invest energy proportional to the electric potential $E_{eq}$ to counterbalance the diffusive force originating from the concentration gradient. When $[K^+]_{ex}$ changes, the diffusive force changes, and so does the energy necessary to maintain the $[K^+]_{in}$. We hypothesize that the information on such changes propagates through the cytoskeleton networks to various cell organelles which then respond appropriately.

The energy spent by the cell to counteract the change in $[K^+]_{ex}$ can be approximated by Nernst Equation. Consider the scenario illustrated in Figure \ref{fig:fig1}a where the external concentration of cell with $[K^+]_{in}$ suddenly changes from $[K^+]_{ex1}$ in normal conditions to $[K^+]_{ex2}$ in stressed conditions. Then, the change in energy required, which is proportional to the change in equilibrium electric potential, can be estimated as,
\begin{equation}\label{eq:eqpot}
    \Delta E_{eq} \equiv E_{eq,2}-E_{eq,1} = \frac{RT}{zF}\ln\frac{[K^+]_{ex2}}{[K^+]_{ex1}}
\end{equation}
where $R$, $T$, $z$, and $F$ are gas constant, temperature, number of elementary charges, and Faraday’s constant, respectively. Note that without loss of generality, we have illustrated in Figure \ref{fig:fig1}a, the scenario where $[K^+]_{ex2}>[K^+]_{ex1}$; however, Eq. \ref{eq:eqpot} should hold in the reverse case too. The information signal that flows from the cell membrane to various cell organelles through random networks of conducting cytoskeleton will scale with potential changes of Eq. \ref{eq:eqpot}. 

To represent the proposed system dynamics, we utilize a 3D parametric cell model based on cytoskeleton’s electrical properties as a mechanism of information signaling as depicted in Figure \ref{fig:fig1}b. Cytoskeletons collectively include actin filaments, intermediate filaments, and microtubules, that range from 10-100 nm in diameter and 1-10$\mu$m in length \cite{58}. Cytoskeleton elements are polyelectrolyte macromolecules that are permanently charged in nature \cite{37, 42, 59}. Based on earlier empirical observations \cite{18,20,30,35,38,39,60,61,62} and computational models \cite{25,33,63,64}, microtubules (MTs) and  Actin fibers (Afs) have the following properties:The hydration shell around tubulin can range from approximately 1 nm to several nms under reduced ionic concentrations, reduced pH or with increased solvent concentration (e.g. DMSO or glycerol). These findings indicate that the environment may controllably tune electrical and EM properties of tubulin and MTs. The net electric charge of tubulin has been shown to be controllable by the environment such that tubulin, which is normally highly negatively charged (up to 50 negative charges per dimer), can become positively charged when pH is lowered below 5 or when the concentration of DMSO reaches 90\% \cite{65}. Importantly, this positively charged tubulin can still polymerize into MTs. It is important to point out that in ionic solutions these charges will be largely (but completely) screened. Some of the counterions congregating around the outer surface of a microtubule are mobile and contribute to the ionic conductivity along the protein polymers \cite{24}.These results were corroborated by electrophoretic mobility measurements and zeta potential determination (66). Polarizability and the index of refraction have been precisely measured and computed as a function of pH with values that vary in the range (4 to 9)$\times$ 10$^{-34}$ Cm$^2$/V, for pH between 6.6 and 7.4 \cite{60, 67}. This is due to a combination of effects resulting from the application of electric fields to microtubules, namely a reorientation of the tubulin’s permanent dipole moment, induced dipole moment generation and ionic bilayer formation around the protein surface. The dielectric constant of tubulin in the same range of pH varies  between 2 and 4, which is much lower than previously reported in the literature and causes strong electric field transmission \cite{18} across the protein in its dimeric and polymeric forms \cite{60}. Both real and imaginary parts of the impedance of MTs and their ensembles were measured at specific concentrations of tubulin ranging from 0.222 $\mu$M to 22.2 $\mu$M and ionic concentrations between 4 and 120 nM \cite{67}. At low ionic concentrations, the Debye length around MTs can reach tens of nm and hence MTs behave as good ionic conductors with a conductivity up to 2-3 orders of magnitude greater than that of the surrounding (cytoplasmic) solution \cite{68}. Conversely, at high ionic concentrations (at or above physiological levels), MTs have Debye lengths on the order of 1 nm and behave as capacitors (charge storage devices for positive counter ions, mainly potassium) while simultaneously exhbiting conductive properties comparable to those of the ionic solution in which they are bathed \cite{60,69}. The cross-over point for the conductive-to-capacitive property of MTs in comparison to the electrolyetic solution is close to 100 mM, which is within the physiological range. This importantly indicates that in cellular environments, the roles of MTs can drastically change depending on the ionic state of the cell. Hence, as electrical circuit elements, MTs can be represented as a resistor-capacitor combination ($R-C$) \cite{21} and in some cases, when the ionic currents can flow along a helical pathway wrapped around the MT cylinder, there is also the inductive component ($L$). Importantly, these electrical characteristics of MTs lead to non-trivial electromagnetic frequency- and ionic concentration-dependent nonlinear response to external stimuli rendering them extremely versatile electromagnetic devices \cite{61}. MTs  provide a range of physical interactions with ions and ionic currents, namely: (a) form a physical barrier that impedes ionic conductance between the electrodes, (b) attract and accelerate ions in a favorable direction leading to increased conductance, (c) condense ions on their surface leading to a reduction in mobile charge carriers in the medium, (d) use C-termini for ion fluxes in and out of the lumen potentially providing helical conduction pathways that can generate solenoid-like \cite{64} behavior of MTs and even memristive properties \cite{70,71} under special conditions of electric field oscillations at fixed frequencies \cite{72}.

Along with conducting electrical and ionic current, a large and persistent change in local electrical energy landscape can result in reassembly of cytoskeleton network as illustrated in Figure \ref{fig:fig1}c. If we assume an ion redistribution in a local region of cytoplasm as a result of an external perturbation, we let $\mathbf{E}$ be the electric field induced in the cytoplasm due ion redistribution. Let $\mathbf{p}=q\mathbf{d}$ be the local cytoskeleton dipole moment in the vicinity of the ion cloud. The deformable cytoskeleton then feels a torque, $\mathbf{r}=\mathbf{p}\times \mathbf{E}$,  and re-aligns until it reaches a stable energy state. We note, however, electrochemical and thermal energy landscape around cytoskeleton is a complex function of the dynamically varying ion concentration making it difficult to model self-assembly of cytoskeletons based solely on only physical laws \cite{73}. 

Electrically conductive cytoskeleton is known to transmit electrical/ionic current signals and in some cases the signals travels in the form of soliton which ideally maintains its shape and propagates with a constant velocity. \cite{18,20,31,32,43,44,74}. Thus, information signals can potentially travel as solitons through the cytoskeleton network and reach various cell organelles. 

Cytoplasmic ion redistribution changes the local electrical potential landscape generating a potential gradient for electric signal flow. However, the local electric potential landscape can be a complex function and current-voltage characteristics are affected by cytoskeleton electrical properties of capacitance and inductance. Here, we follow a parametric approach and define a graphical model of 3D spherical cell (Cell-Reservoir) in which the cytoskeleton components are viewed as a network of effective resistors that conduct electric signals following Ohm’s law, $I_k  = G_k \Delta V$, where $G_k$ is the effective conductance or inverse impedance of the $k^{th}$ resistor and $\Delta V$ is the potential difference across the resistor. The information source originates at the peripheral cytoplasm, and, for simplicity, we assume that the potential distribution follows $\sim \exp(-kr)/r$ law, where $1/k$ is the Debye length. The signal will then flow directionally from higher to lower potentials. These information signals, which are electrical in nature, can aggregate at the cytoskeleton crossover points or multiply at split points as shown in Figure \ref{fig:fig1}d following Kirchoff’s current law. Signals propagating through the cytoskeleton first reaches the organelles closest to the source and subsequently to other organelles, distance-wise. Depending on the type of signals, cell organelles initiate appropriate biological processes that travel diffusively. An animation of information flow from a point source at the peripheral cytoplasm is shown in Figure \ref{fig:fig1}e.

The cytoskeleton ionic conductivity changes with the cytoplasmic ionic concentration. For instance, microtubules ionic conductance was reported to be highest in the presence of monovalent K$^+$ and Na$^+$ in the background solution, which reduced for divalent and trivalent ions. To account for the intrinsic variability, we assign a log-normal distributed conductance to the cytoskeleton. Several experiments have yielded a range of cytoskeleton conductivity: 0.124 S/m for actin filament \cite{59}, and 0.15 S/m  \cite{60} to 0.25S/m \cite{62} for microtubules. As noted above, these conductivity values are strongly dependent on the ionic concentration. Considering this variability in conductivity due to background ionic concentration and bundling of cytoskeleton filaments, we assume a range of 0.1-100 S/m. The conductance of a filament of length 10$\mu$m and cross-section of 1$\mu$m$\times$1$\mu$ m is roughly $10^{-8}-10^{-5}$ S ($G=\sigma A/l$). Note that we assume here the cross section of a filament bundle since individual MTs have an outer diameter of 25 nm. Then, assuming a range of change in ionic concentration from 1-100 mM at body temperature, Eq. \ref{eq:eqpot} yields a potential range of, 0.01-0.1 V, which corresponds to signal strength in order of  $10^{-10}-10^{-6}$ A. A similar current value has been reported for cytoskeletons modeled as memristors\cite{70}. 

\begin{figure*}
    \centering
    \includegraphics[width=0.8\textwidth]{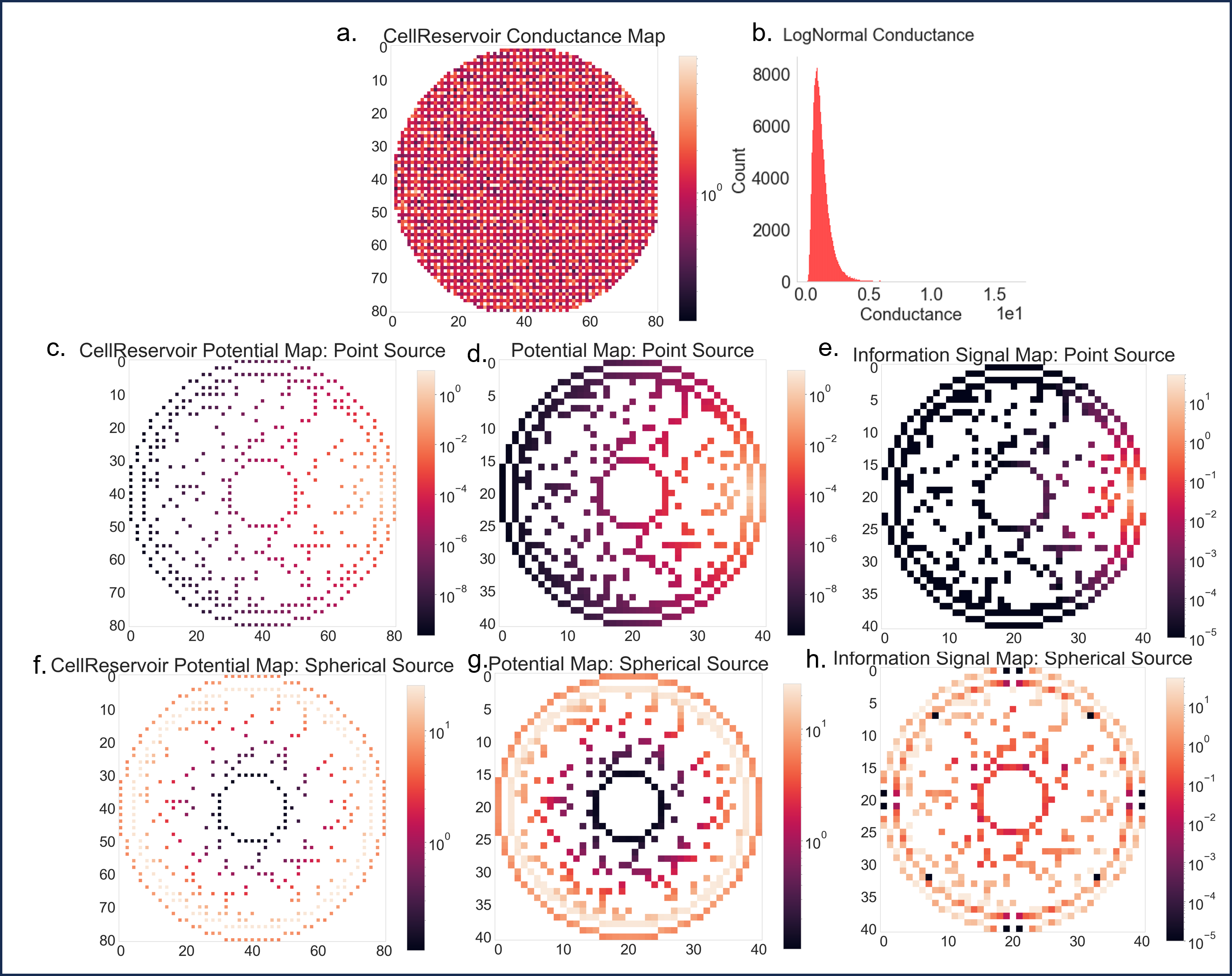}
    \caption{\small \textbf{Cell-Reservoir Electrical Properties:} \textbf{a.} 2D cross-sectional conductance map of Cell-Reservoir edges, which represents the connection strength between two neighboring voxels. \textbf{b.} A log-normally distributed conductance was assigned to the Cell-Reservoir edges. \textbf{c-h} are Cell-Reservoir potential map, Cell potential map, and Cell information signal map for point source and spherical source respectively. Potential distribution follows the $\exp(-kr)/r$ law and the information flow follow Ohm’s and Kirchhoff’s current laws.}
    \label{fig:fig3}
\end{figure*}

\subsection{Cell-Reservoir: A quasi-physical reservoir model for cellular information dynamics}
Cell-Reservoir is a quasi-physical model that fuses prior knowledge on bioelectrochemistry into a data-driven approach. Such hybrid framework provides an algorithmic approach for modeling the dynamics of sub-cellular processes that is capable of learning spatiotemporally resolved cellular behavior. Computationally, Cell-Reservoir is a graph-based reservoir system connected to a decision-making module that is capable of learning cellular decision-making process from measurement data.  The model descriptions are as follows.  

\subsubsection{Cell Geometry}
We consider a voxel-based 3D spherical cell consisting of three closed surfaces that represent cell membrane (CM), peripheral cytoplasm (PC) and central organelles (CO) (nucleus and the surrounding endoplasmic reticulum) as shown in Figure \ref{fig:fig2}.  For simplicity, we consider concentric surfaces with radius $r_{CM} = 10\mu$m,  $r_{PC} = 9\mu$m, and $r_{CO}=2.5\mu$m. The cytoplasm is filled with randomly generated network channels representing cytoskeleton. For this work, we filled 20\% of cytoplasm with cytoskeletons. Further analysis with different volumes is presented in the Results section. Since it is difficult to visualize solid 3D objects in 2D, we have presented a UMAP projection in Figure \ref{fig:fig2}c and in Supplementary Materials (SM) Section \ref{sec:sm1}. We have included three rows corresponding to projections on same cell with all components, without cell membrane, and without cell membrane and peripheral cytoplasm respectively. The left column shows the voxels while the right column shows the UMAP nearest neighbor (NN) connectivity map.

\subsubsection{Data Structure}
Cell-Reservoir is a grid graph $G(V,E)$ on a 3D space composed of vertices ($V$) and edges ($E$) as shown in Figure \ref{fig:fig2}d. The grid is indexed with a 3D indexing system ($i,j,k$), which corresponds to $x$, $y$ and $z$ direction, respectively. For discretizing, a space of volume $\Lambda$ is divided into $n\times n\times n$ ($n^3$) voxels of unit volume $\Lambda/n^3$. For representing spatially correct edges we select a 3D array of size $|G| = 2n-1\times2n-1\times2n-1$ ($\sim 8n^3$) and assign vertices to the even number indices ($2i,2j,2k$). Each vertex represents one voxel, and one vertex is surrounded by 26 edges ($d^3-1$). There are $7n^3-12n^2-3n-1$ ($\sim7n^3$) edges in Cell-Reservoir. Computationally, by exploiting the spatial information of the edges, grid graph approach provides an efficient alternative to adjacency matrix approach, which is a 2D, $n^3\times n^3$ ($n^6$) matrix. 

\subsubsection{Information Dynamics}
We follow iterative methods for modeling information dynamics. The central assumption is that at every time iteration, information signals can travel one voxel. Information dynamics essentially follow Huygens’ principle in which a source voxel transmits information signal to its nearest neighbors via 26 surrounding edges, and every receiving voxel then becomes a source of information in the next time step. Duration of a time iteration depends on the size of a voxel: small unit volume results in small time step. Cell-Reservoir keeps track of information boundary at each time iteration; as information travels information boundary covers more volume. 

\subsubsection{R-Ball}
Cell-Reservoir stores edge parameters in its spatially correct location. We exploit this property by defining a R-Ball function of size  5$\times$5$\times$5 that can traverse through the Cell-Reservoir for updating information flow and tracking the information boundary. R-ball can only be centered around a vertex or even number indices. R-ball essentially keeps track of the nearest neighbors.

\begin{figure*}
    \centering
    \includegraphics[width=0.9\textwidth]{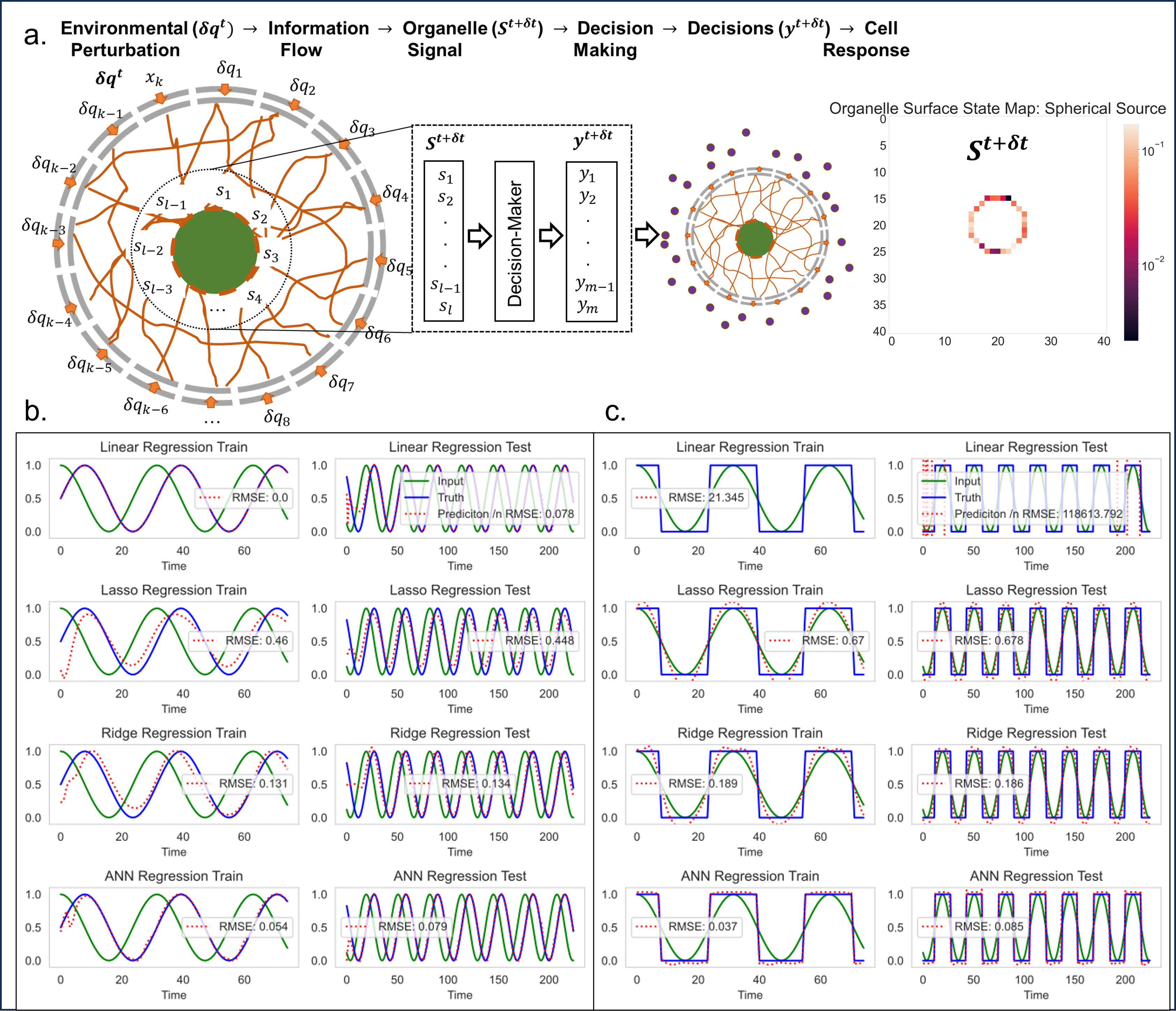}
    \caption{\small \textbf{Reservoir Computing for Cellular Decision-Making}. \textbf{a.} Cell-Reservoir has a decision-making (Readout) layer for learning various intracellular decision-making tasks.  Information on environmental perturbation, $\delta q^t$, at time $t$ is propagated through the cytoskeleton which reaches the organelle surface node at time $t+\delta t$. The organelle memory state signal, $S^{t+\delta t}$, also shown for a spherical source, can be fed into a Readout Layer for decision-making. The cell decisions, $y^{t+\delta t}$ is then used for appropriate cell response. \textbf{b.} Cell-Reservoir learning sine wave response and \textbf{c.} square wave response for a cosine wave input via linear, lasso, ridge, and artificial neural network readout layer. The left columns show the training mode, and the right columns show the testing mode. Each figure includes the root mean square error between the ground truth and prediction value.}
    \label{fig:fig4}
\end{figure*}

\subsubsection{Electrical Properties}
Figure \ref{fig:fig3} summaries Cell-Reservoir electrical properties of $81\times81\times81$ Cell-Reservoir with unit voxel volume of 0.5$\mu$m$\times$0.5$\mu$m$\times$0.5$\mu$m. To represent variability in cytoskeleton conductance, we assigned edges with a log-normally distributed conductance of mean 0.1 units and deviation of 0.5 units as shown in Figure \ref{fig:fig3}a and \ref{fig:fig3}b. In this work, we considered two end cases: point source and spherical source, which were placed at the peripheral cytoplasm. Assuming $q\exp(-kr)/r$ law with a typical value of Debye length ($1/k$) of 1$\mu$m, we first found the potential distribution for electric charge ($q$) of 1 unit, as shown in Figures \ref{fig:fig3}c and \ref{fig:fig3}d for point source and in Figures \ref{fig:fig3}f and \ref{fig:fig3}g for spherical source. Then we found the information dynamics iteratively where information flows from one voxel to its neighboring voxel following Ohms and Kirchhoff current law, as shown in Figure \ref{fig:fig3}e and \ref{fig:fig3}h for point source and spherical source respectively.

\begin{figure*}
    \centering
    \includegraphics[width=0.9\textwidth]{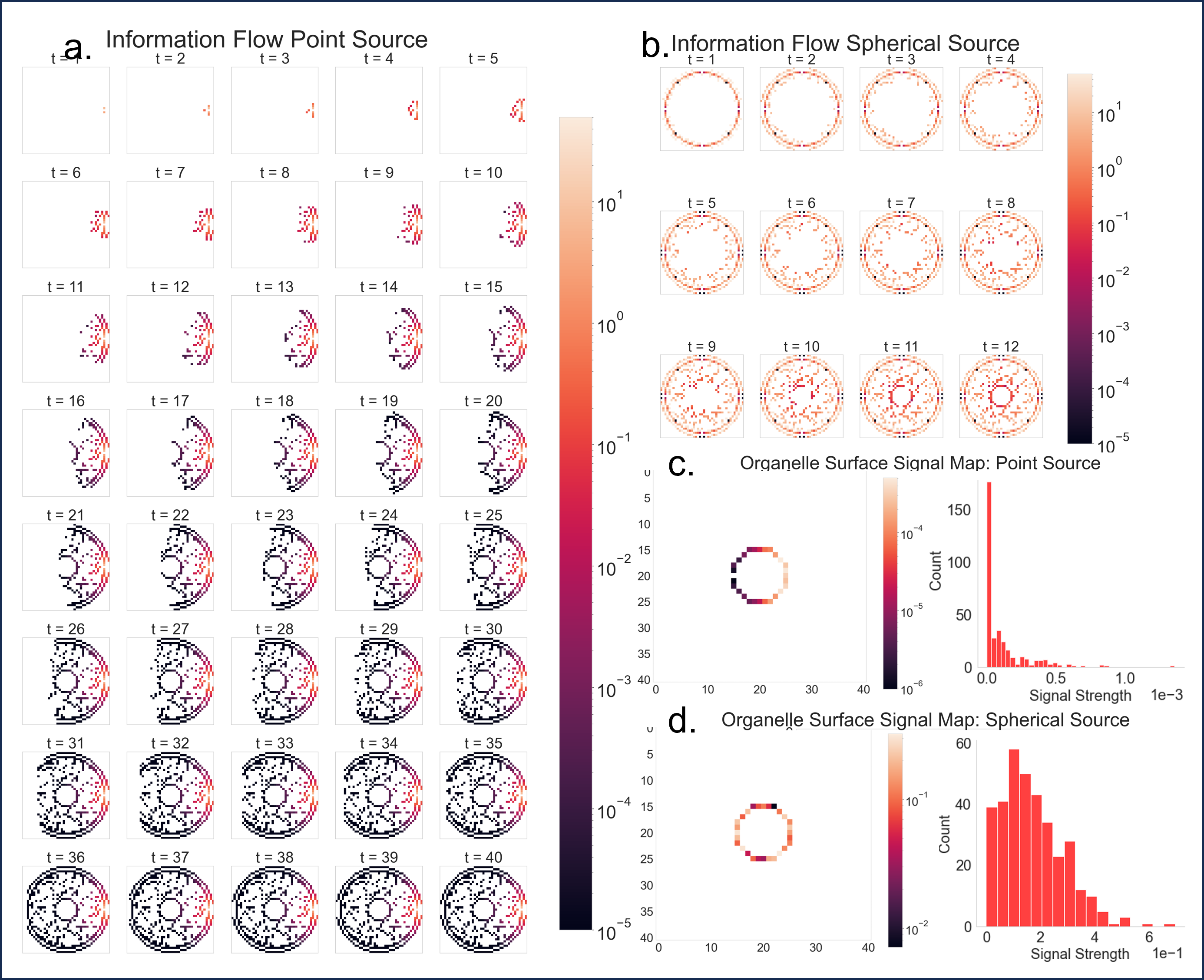}
    \caption{\small \textbf{Spatiotemporally resolved Intracellular Information Dynamics}. Snapshots of information flow in a $81\times 81\times 81$ Cell-Reservoir or a $41\times 41\times 41$ Cell for \textbf{a.} Point Source for 40 time-steps (each time steps is about $0.05-0.5 \mu s$) and \textbf{b.} Spherical Source for 12-time steps, color coded with signal strength. Central organelle surface signal map and signal histogram for \textbf{c.} point source and \textbf{d.} spherical source.}
    \label{fig:fig5}
\end{figure*}

\subsubsection{Decision-Making}
The Cell-Reservoir model can be utilized for modeling an organelle’s decision-making task (or response to a stimuli) via an RC framework as shown in Figure \ref{fig:fig4}a. The task can be as simple as a binary task, e.g., cell organelles activated or not activated, or as complex as decision-making based on a time sequence of events. Given a pair of input signal ($X$) and cell response ($y$), Cell-Reservoir can learn to make optimal decisions for appropriate cell response via a Readout Layer using a regression approach, which minimizes a specified cost function such as mean square error (MSE), $L(\hat{y},y)=\frac{1}{n} \sum_{i=1}^{n}(y_i-\hat{y}_i)^2$ , for model predictions $\hat{y}$. The details of Cell-Reservoir RC framework are as follows.

Consider an environmental perturbation of strength $\delta q^t (X^t)$ in the vicinity of cell membrane. The Cell-Reservoir potential distribution due to the perturbation is then approximated as,
\begin{equation}\label{eq:pot}
V_i^t = \sum_s \delta q_s^t  \frac{\exp(-kr_{is})}{r_{is}},
\end{equation}
where $r_{is}$ is the Euclidean distance between vertex $i$ and source vertex $s$. We assume an adiabatic process where the time scale of external perturbation is relatively slower than the time taken by the electric field to permeate through the cell. At every time, $t$, we calculate the potential map of cell graph originating from environmental perturbation $\delta q^t$. This perturbation results in an information signal that originates at the peripheral cytoplasm and travels through the labyrinth of cytoskeleton. As shown in Figure \ref{fig:fig2}d, Cell-Reservoir stores two values for every occupied vertex $i$: (1) physical signal,
\begin{equation}\label{eq:cur}
 I_i=\sum_{n\in NN_{i}} i_n = \sum_{n\in NN_i} G_{in} (V_i-V_n),  
\end{equation}
where $NN$ is the nearest neighbor occupied vertex of vertex $i$, $i_n$ is the current contribution from $NN_n$, $V$ is the vertex potential, and $G_{in}$ is the conductance of the edge connecting vertex $i$ and $n$; and (2) memory state,
\begin{equation}\label{eq:mem}
S_i^t=\beta S_i^{t-1} + (1-\beta) \tanh(I_i),
\end{equation}
where $\beta$ is the memory retention rate and $0<\beta<1$. $\beta=0$ corresponds to no memory whereas $\beta=1$ corresponds to no learning. At $t=0$ we assume that $S_i^0=0$. The hyperbolic tangent activation function squashes the physical signal in between -1 and 1. The information signal takes finite time $\delta t$ to reach to the cell organelle. The memory signals from the surface of the central organelle, $S^{t+\delta t}$ is then fed into the decision-maker (ReadOut Layer) which yields decision state $y^{t+\delta t}$. The cell then responds according the to decision state. Figure \ref{fig:fig4}b and \ref{fig:fig4}c shows an example of RC learning for spherical sources. We trained the proposed Cell-Reservoir to learn a Sine wave response and square wave response originating from a cosine wave input signal. Sine wave response can be interpreted as a periodic response to periodic environmental perturbation with a delay and square response can be interpreted as a binary response (threshold) to periodic environmental perturbation. 

We first trained a Cell-Reservoir with four different ReadOut Layers: Linear, Lasso, Ridge, Artificial Neural Network (ANN) regression. Note that all four ReadOut layers learned the sine response well, while only ANN learned the square wave well. Since ANN are non-linear models, such behavior is not unexpected. Additionally, bias effect can be seen to affect learning during initial time steps, a typical behavior of exponential moving averages. During initial learning steps, the cell has no memory which biases memory states of Eq. \ref{eq:mem} toward 0. Various bias correction methods exist such as dividing by $1-\beta^t$ or adding a warmup buffer and simply discarding the first n states, however we added no such correction as we find such behavior natural to living beings which learns to be efficient over time. 

\begin{figure*}
    \centering
    \includegraphics[width=0.7\textwidth]{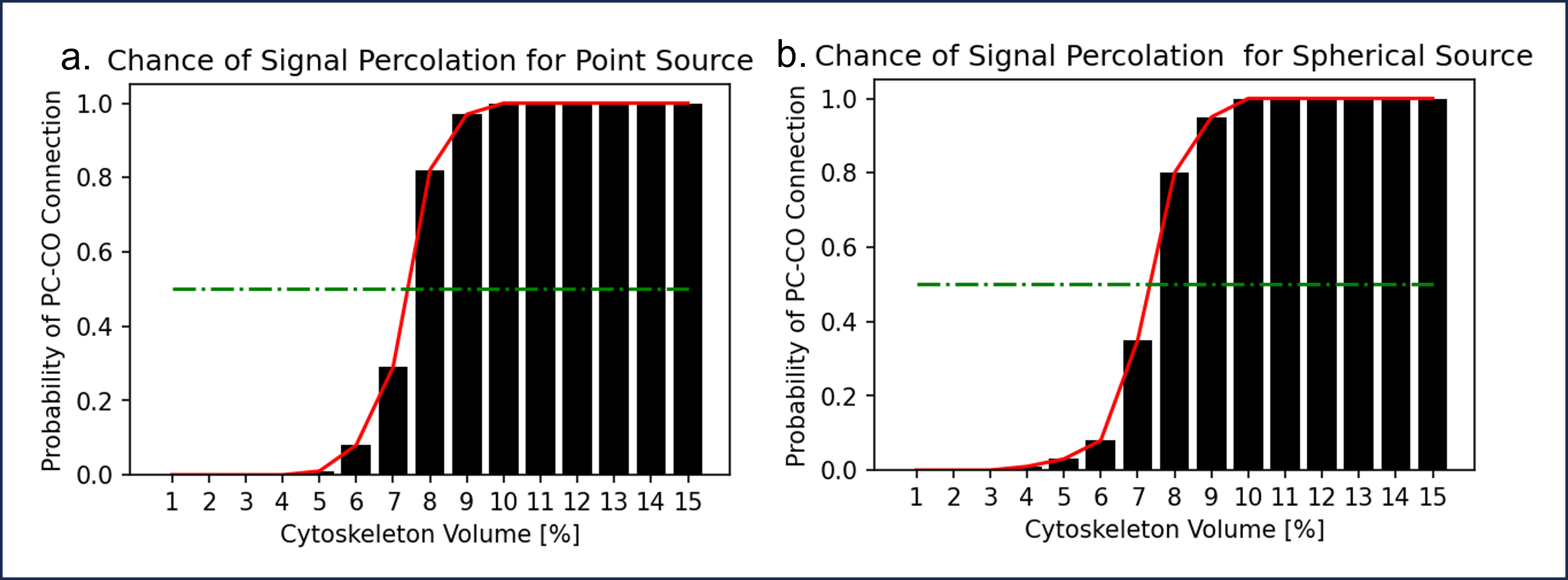}
    \caption{\small \textbf{Percolation Analysis}. The chance of signal percolation from peripheral cytoplasm (PC) to central organelle (CO) as a function of cytoskeleton volume per cytoplasm for \textbf{a.} point source and \textbf{b.} spherical source obtained from repeating 100 trials per cytoskeleton volume. Each trial randomly generated cytoskeletons from uniform distribution and checked if CO can receive any information originating from the PC.}
    \label{fig:fig6}
\end{figure*}
\begin{figure*}
    \centering
    \includegraphics[width=0.8\textwidth]{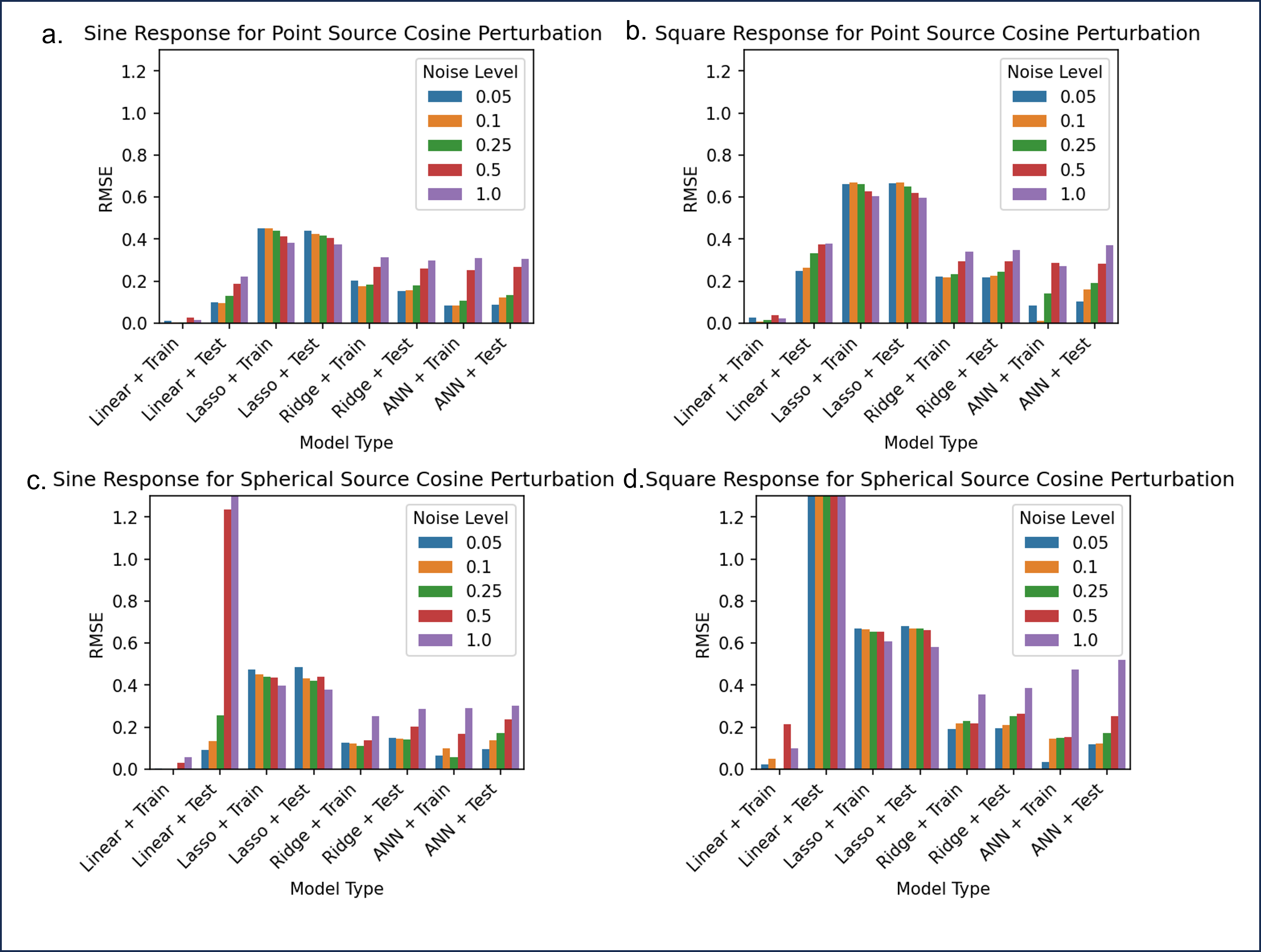}
    \caption{\small \textbf{Noise Analysis}. Cell-Reservoir’s robustness to noise analyzed for cosine perturbation and \textbf{a.}, \textbf{c.} sine and \textbf{b.}, \textbf{d.} square response for point source and spherical source, respectively with Linear, Lasso, Ridge, and Artificial Neural Network (ANN) as decision-makers. Five levels (standard deviation) of normally distributed noise were added to cosine wave and the Cell-Reservoir were trained in the first 1/4$^{th}$ of the noisy signal and tested on the last 3/4$^{th}$. Root Mean Square Error (RMSE) between the predicted response and ground truth is reported.}
    \label{fig:fig7}
\end{figure*}

\section{Results}
\subsection{Cell-Reservoir’ spatiotemporal structure captures directed nature of information dynamics}
Figure \ref{fig:fig5} illustrates the spatiotemporal capabilities of Cell-Reservoir and directed nature of information dynamics in a 81$\times$81$\times$81 Cell-Reservoir with 0.5$\mu$m$\times$0.5$\mu$m$\times$0.5$\mu$m voxels for point source and spherical source. In the case of a point source, which was placed at the right end, the information can be seen flowing towards the left, covering more volume at each time step. While for the spherical source, the information can be seen traveling towards the center. In either case, information travels from voxels at higher potential to the voxels at lower potential, like fluid travelling from higher to lower grounds. As expected, the signal from spherical source took less time ($\sim$12 steps) to saturate the whole cell in comparison to a point source ($\sim$40 steps). However, in both cases, the information took about the same time ($\sim$10 time steps) to reach the central organelle, independent of the source distribution. 

For both cases, a unit charge was applied as the source strength, thus, Figure \ref{fig:fig5} naturally shows a much higher signal strength for spherical source due to larger aggregate charge. From the histograms of Figures \ref{fig:fig5}c and \ref{fig:fig5}d, we can see that the organelle signal distribution is different due to spatially different sources.

Considering the fact that cytoskeletons are similar to nerve fiber, we assume that the intracellular information travels at a speed very similar to that of nerve conduction velocity, which is highly sensitive to nerve diameter among other factors and can range from 1 m/s to 100 m/s.(75, 76) Nerves with smaller diameters ($\sim$ 1$\mu$m) conduct electrochemical impulses at slower speed. Thus, for a cytoskeleton filament bundle of 1 $\mu$m, we assume that the information can travel at 1-10 $m/s$ speed and estimate that intracellular information takes about 2-20 $\mu$s to traverse a cell of radius 10 $\mu$m, and about 0.05-0.5 $\mu$s to traverse one voxel of length 0.5 $\mu$m. This means that information on external perturbation takes about  0.5-5 $\mu$s (10-time steps) to reach to the central organelles and at least the same time to respond to any external perturbation. 

\subsection{Cytoskeleton volume has threshold for Signal Percolation}
Figure \ref{fig:fig6} summarizes the results from percolation analysis on the Cell-Reservoir carried out for point source and spherical source to determine the minimum cytoskeleton volume needed for information signal originating from the peripheral cytoplasm to successfully percolate to central organelle. We repeated 100 trials for each cytoskeleton volume ranging from 1\% to 15\% of the cytoplasm volume sandwiched by peripheral cytoplasm and central organelle. Altogether, a total of 3000 trials were carried out between the point and spherical sources. In each trial, a random configuration of cytoskeleton was generated following uniform distribution, information dynamics were simulated until saturation, and whether information signal reached the central organelle was assessed. We found that signal percolation was independent of source distribution, which is expected considering Huygen’s principle-like information dynamics process. In both cases, we found a step transition from non-percolating to a percolating system in between 5\% to 10\% of the cytoplasmic volume. The volume of about 7.5\% corresponded to 50\% probability of percolation and volume of over 10\% guarantees percolation. UMAP projection and signal map for various cytoskeletons volume are presented in \ref{sec:sm1} and \ref{sec:sm2}.

\subsection{Signal to Noise Analysis}
To evaluate Cell-Reservoir’s robustness to signal noise, we carried out a noise analysis by adding normally distributed random noise, $\mathcal{N}(0,\sigma^2)$, to the input signal during RC learning. Figure \ref{fig:fig7} presents summary statistics for point and spherical source which were perturbed periodically following cosine function for learning sine response and square response. For noise analysis, Cell-Reservoir with four different decision-making layers (Linear, Lasso, Ridge, and ANN) were analyzed on five noise levels ($\sigma$) of strength 0.05, 0.1, 0.25, 0.5, and 1.0 were added to cosine function input of amplitude 1.0. The model was trained on the first 1/4$^{th}$ of the noisy signal and tested on the last 3/4$^{th}$ and a root mean square error (RMSE) was found between the ground truth (sine and square wave) and the prediction. From the numerical experiment, we observed that Linear Layer was the least robust and ``blew up" for the spherical source, and ANN was the most stable decision-maker model. Additionally, Lasso model did not scale with the signal-to-noise ratio. As a result, ANN layers were selected for subsequent experimentation. Additional plots are presented in \ref{sec:sm3}. 

\begin{figure*}
    \centering
    \includegraphics[width=0.9\textwidth]{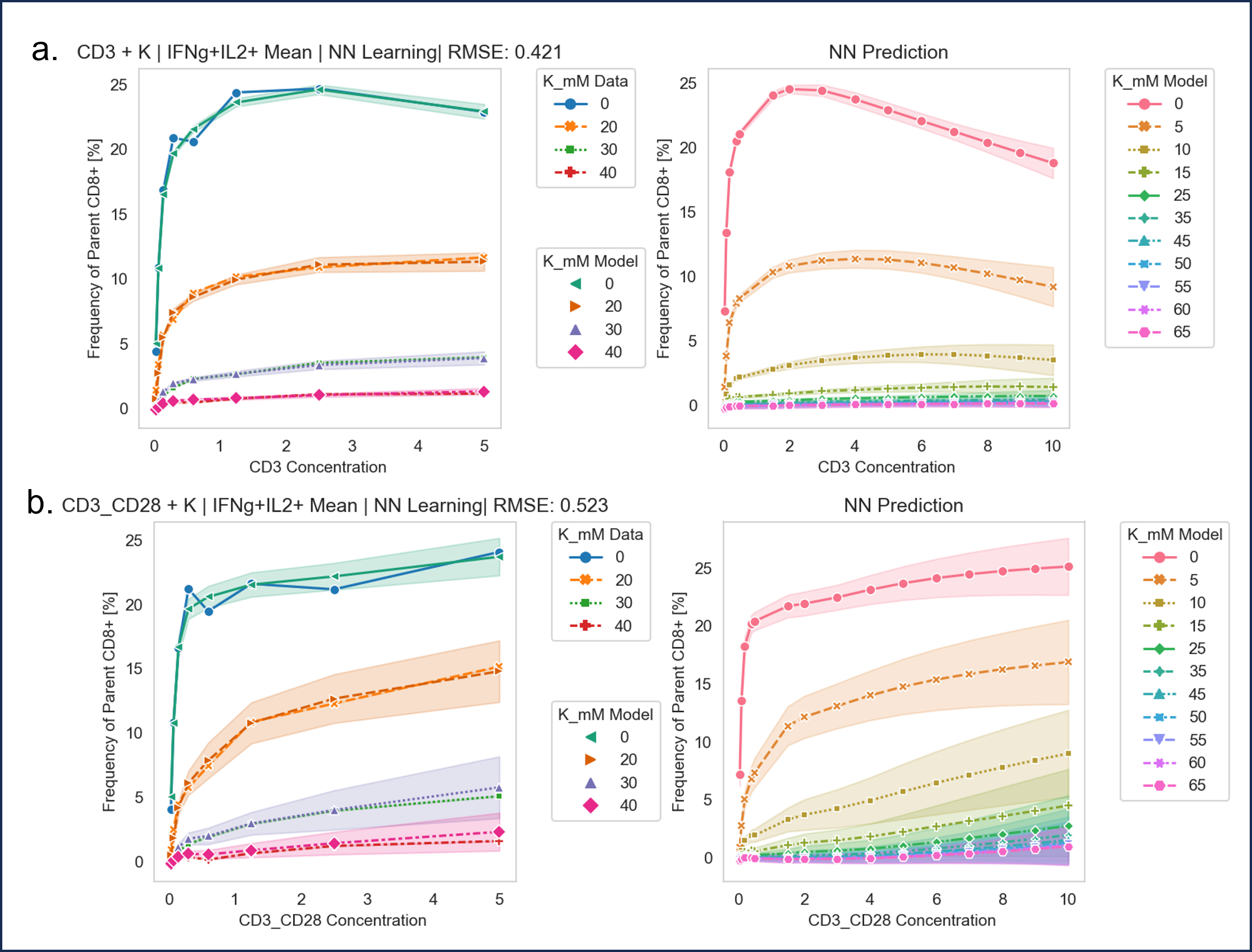}
    \caption{\small \textbf{Experimental Evaluation of Cell-Reservoir on Ionic Immune Suppression Data}. Cell-Reservoir learning to mimic CD8 activated cell behavior in the presence of $[K^+]$ and \textbf{a}. CD3 and \textbf{b.} CD3 + CD28. The left column summarizes training and validation, and the right column summarizes prediction for concentration values in between and beyond the available data. Artificial Neural Network with L1 and L2 norm regularization was applied as the decision-making layer.}
    \label{fig:fig8}
\end{figure*}

\subsection{Experimental Evaluation of Cell-Reservoir on Ionic Immune Suppression Data}
Eil et al. found that intracellular K$^+$  ion released by cellular necrosis within tumor microenvironment elevated extracellular $[K^+]$ which suppressed T cell effector function leading to tumor growth \cite{77}. In this work, we use flow cytometry data from the work of Eil et al. and applied Cell-Reservoir model to CD8 T cells to learn and predict immune suppression behavior as shown in Figure \ref{fig:fig8}. In the experiment CD8 cells were introduced to solutions with various CD3 (0-5 mM) and $K^+$ (0-40 mM) concentrations and a combination of CD3 + CD28 (0-5 mM) and $K^+$ (0-40 mM) concentrations, and the proportion of CD8 cells producing Interferon gamma (IFN-$\gamma$) and Interleukin 2 (IL2) among other cytokines were measured via flow cytometry. IFN-$\gamma$ and IL2 production indicates that a T-cell is activated and is healthy/proliferating. 

We took the following steps to apply Cell-Reservoir to this experimental data. First, from the Flowjo software, we acquired the frequency of parent CD8 cell with positive IFN-$\gamma$ and IL2 production which was interpreted as the probability of a CD8 cell to produce both IFN-$\gamma$ and IL2. Additionally, since the experiments were repeated three times, we considered the mean value. For environmental perturbation, we assumed $\delta q^t=[K^+]_{in}-[K^+]_{ex}$, where $[K^+]_{in}\sim$ 150 mM and interpreted the data corresponding to the four values of $[K^+]_{ex}$ as sequential data starting from 0 mM and rising to 40 mM. In the experiments, the CD8 cells were activated via CD3 and in another experiment they were activated via CD3+CD28. We input this information into the decision-making layer via concatenation—we populated a vector of size equal to the central organelle surface vertex with copy of CD3 concentration value, concatenated with the surface memory state and then fed into the decision-making layer. To increase generalizability, we implemented both L1 and L2 norms (elastic net) regularizations to the ANN. We trained and validated our model as shown in the left columns of Figure \ref{fig:fig8}a. and \ref{fig:fig8}b. We made predictions on both interpolated and extrapolated values as shown in the right columns. We applied statistical ensemble for quantifying model uncertainty by repeating 20 trials for each $[K^+]_{ex}$ value. For the training set, we obtained RMSE value of 0.42\% for CD3 activation and 0.52\% for CD3+CD28 activation. As expected, due to limited sample size, we found the model to be heavily influenced by the available data and the data trend, as can be seen by opposite slope of the $[K^+]=0$ line in between the two types of activation. Nevertheless, we can see that Cell-Reservoir can learn ion-induced cell behavior from available data set.

\section{Summary}
Here, we address the complex dynamics of cellular response to environmental information. Prior studies \cite{1} have explicitly linked the speed and accuracy of cellular acquisition, analysis, and response to environmental information to evolutionary fitness. These dynamics, therefore, are subject to continuous Darwinian optimization. 

Existing models of cellular information dynamics focus on ligand binding followed by signal transduction through molecular pathways, which requires 3-dimensional diffusion of macromolecules. However, computer simulations have demonstrated communication by diffusion through the cytoplasm degrades transmission of temporal and spatial environmental information \cite{78}. Thus, we view these information dynamics as insufficient to produce rapid and spatially directed responses to acute, life-threatening perturbations.

Here, we propose an additional, currently unrecognized, intracellular communication network built upon the backbone of large transmembrane ion gradients observed in all living cells. In this model, external perturbations affect specific gates on membrane ion channels that communicate precise spatial and temporal information via local flow of ions along pre-existing concentration gradients. Rapid information analysis and response is generated by altered functions and localization of peripheral membrane proteins \cite{79} induced by local fluctuations of cytoplasmic ion concentrations. If the response is sufficient, membrane ion pumps restore the original gradient. 

However, larger perturbations that produce prolonged and/or large spatial scale changes in cytoplasmic ion concentration require regional or global cellular response. Within our model, these perturbations are communicated through rapid ion-induced self-assembly of cytoskeleton. Multiple studies have documented ions and electrons can flow along microtubules and microfilaments allowing wire-like transmission of information. The dual functions of information transmission and biomechanical transport of the cytoskeleton can engage the endoplasmic reticulum and mitochondria which are deeply enmeshed within it \cite{26,56,80}. This allows rapid transport of these organelles to perturbation to produce the necessary energy and macromolecules for optimal response. In addition, a more global response can be generated by direct connections of the cytoskeleton with the nuclear membrane via the KASH complex which directly links the genome with the cytoskeleton and other components of the cell \cite{29}.  

Here we consider a simplified one source - one target model of these complex dynamics. Simulations within our Cell-Reservoir modeled information transmission and response network demonstrate it can rapidly communicate information from the membrane to intracellular organelles permitting an optimal cellular response. We demonstrate experimental observations of cellular changes in response to altered environmental K$^+$ changes caused by cancer cell death are consistent with the model predictions. 
Our framework can be extended to multi-source multi target problems by summing potential contributions from multiple sources. This model produces predictions that can be tested in subsequent empirical studies.

\section*{Acknowledgement} 
We thank Dr. Robert Eil (Oregon Health \& Science University) for providing flow cytometry data for this work. DN acknowledges Dr. Prabesh Khatiwada (Columbia University) for insightful discussion on cellular immune response and flow cytometry. 

\section*{Code Availability}
Source codes and data are available in \href{https://github.com/DipeshNiraula/Intracellular-Information-Dynamics}{Github}.

\onecolumn
\renewcommand{\appendixname}{Supplementary Materials}
\appendix
\renewcommand{\thesection}{SM\arabic{section}}
\begin{center}
 {\huge \textbf{Supplementary Materials}}
\end{center}

\section{UMAP of 3D Cell-Reservoir for Various Cytoskeleton Volume \label{sec:sm1}}
\subsection{1\% Cytoskeleton Volume}
\begin{center} 
\includegraphics[width=0.8\textwidth]{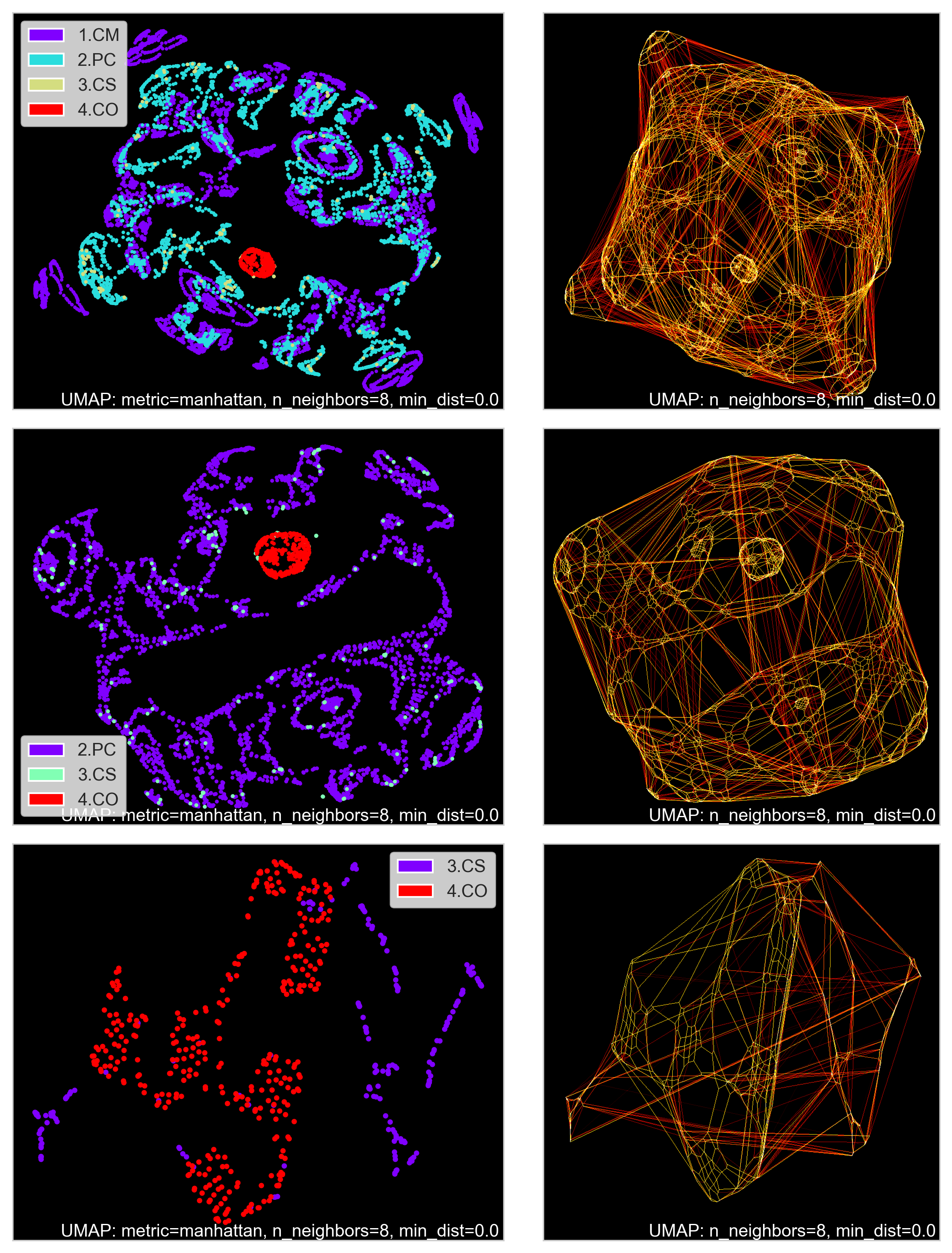}
\end{center}

\subsection{5\% Cytoskeleton Volume}
\begin{center} 
\includegraphics[width=0.8\textwidth]{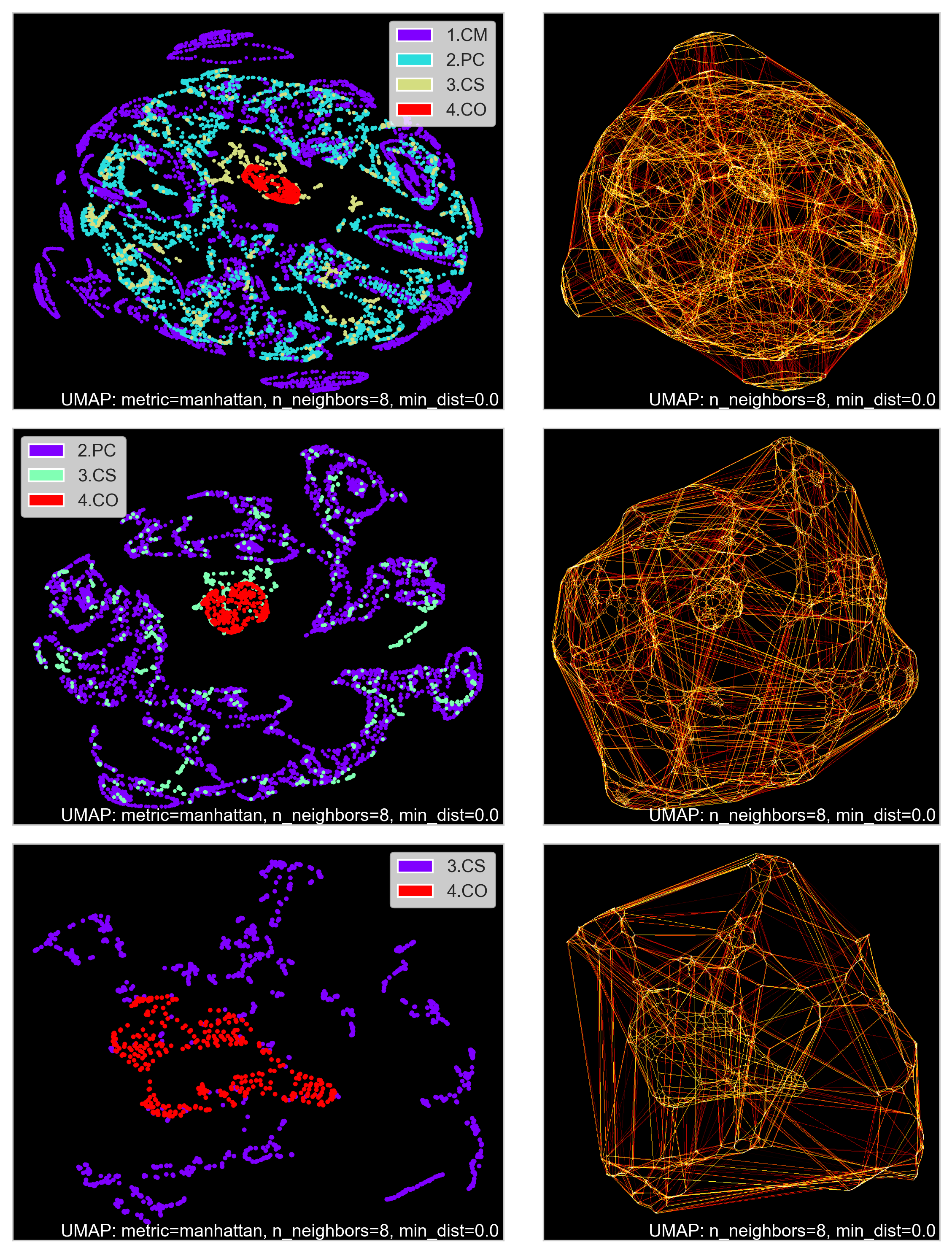}
\end{center}

\subsection{7\% Cytoskeleton Volume}
\begin{center} 
\includegraphics[width=0.8\textwidth]{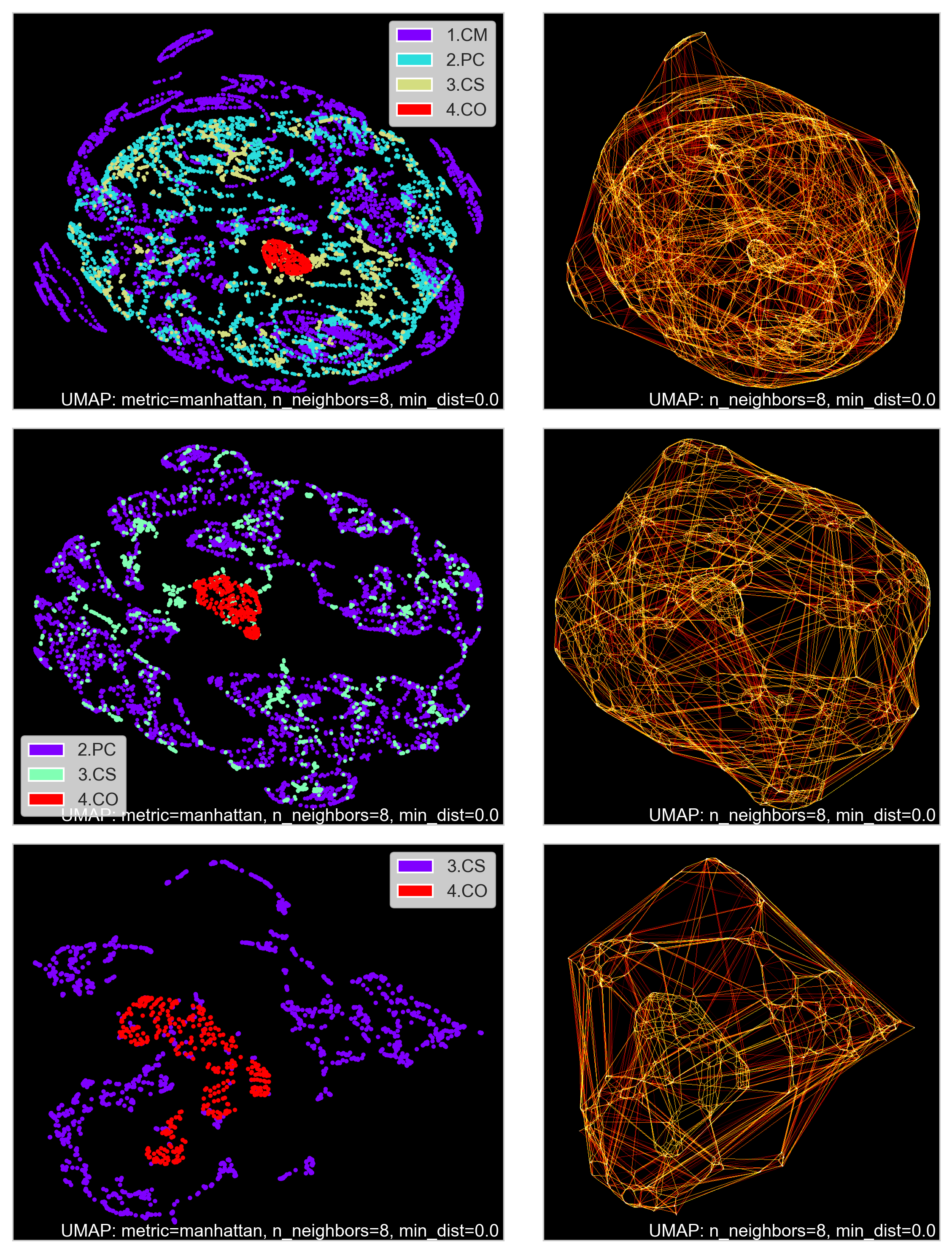}
\end{center}

\subsection{8\% Cytoskeleton Volume}
\begin{center} 
\includegraphics[width=0.8\textwidth]{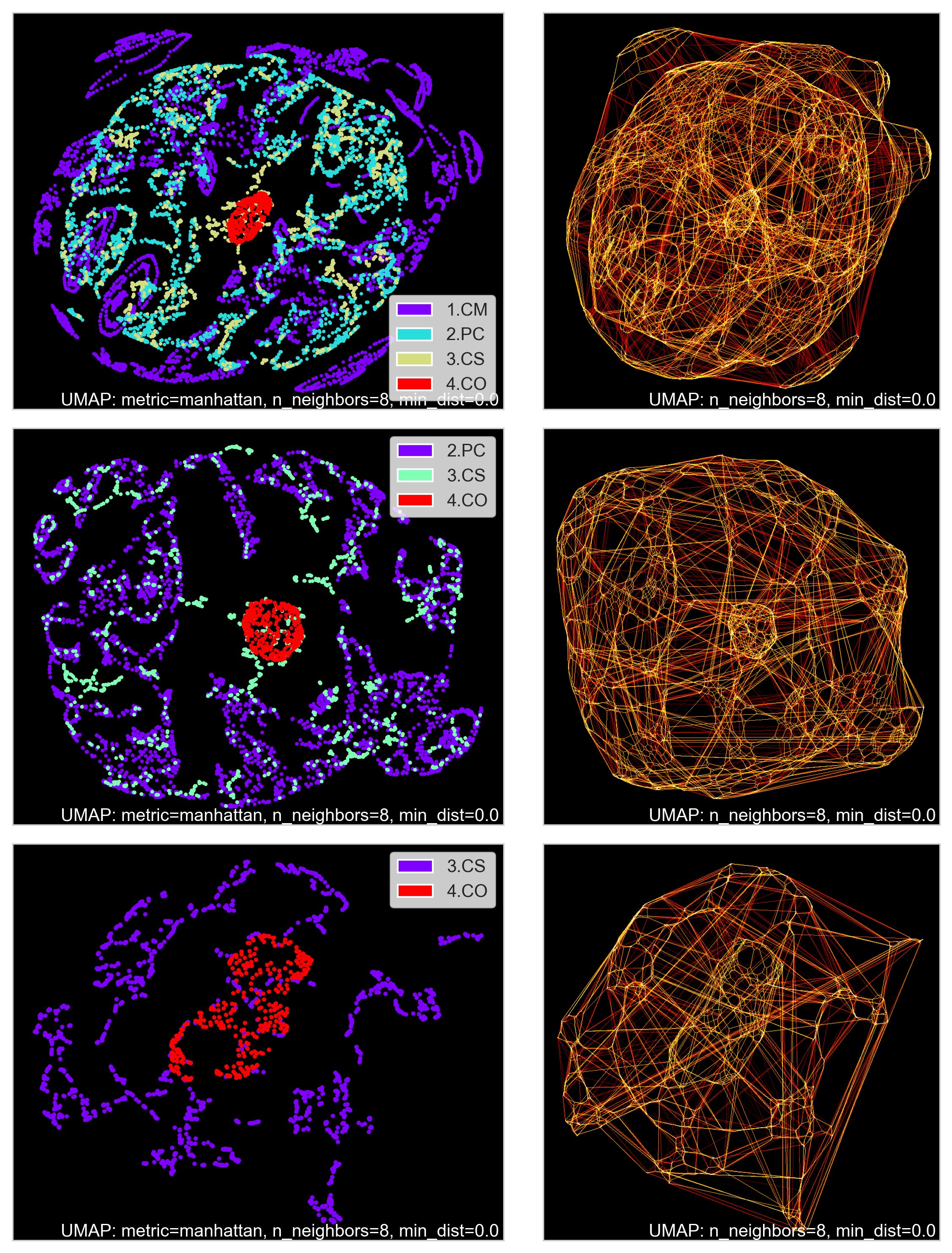}
\end{center}

\subsection{10\% Cytoskeleton Volume}
\begin{center} 
\includegraphics[width=0.8\textwidth]{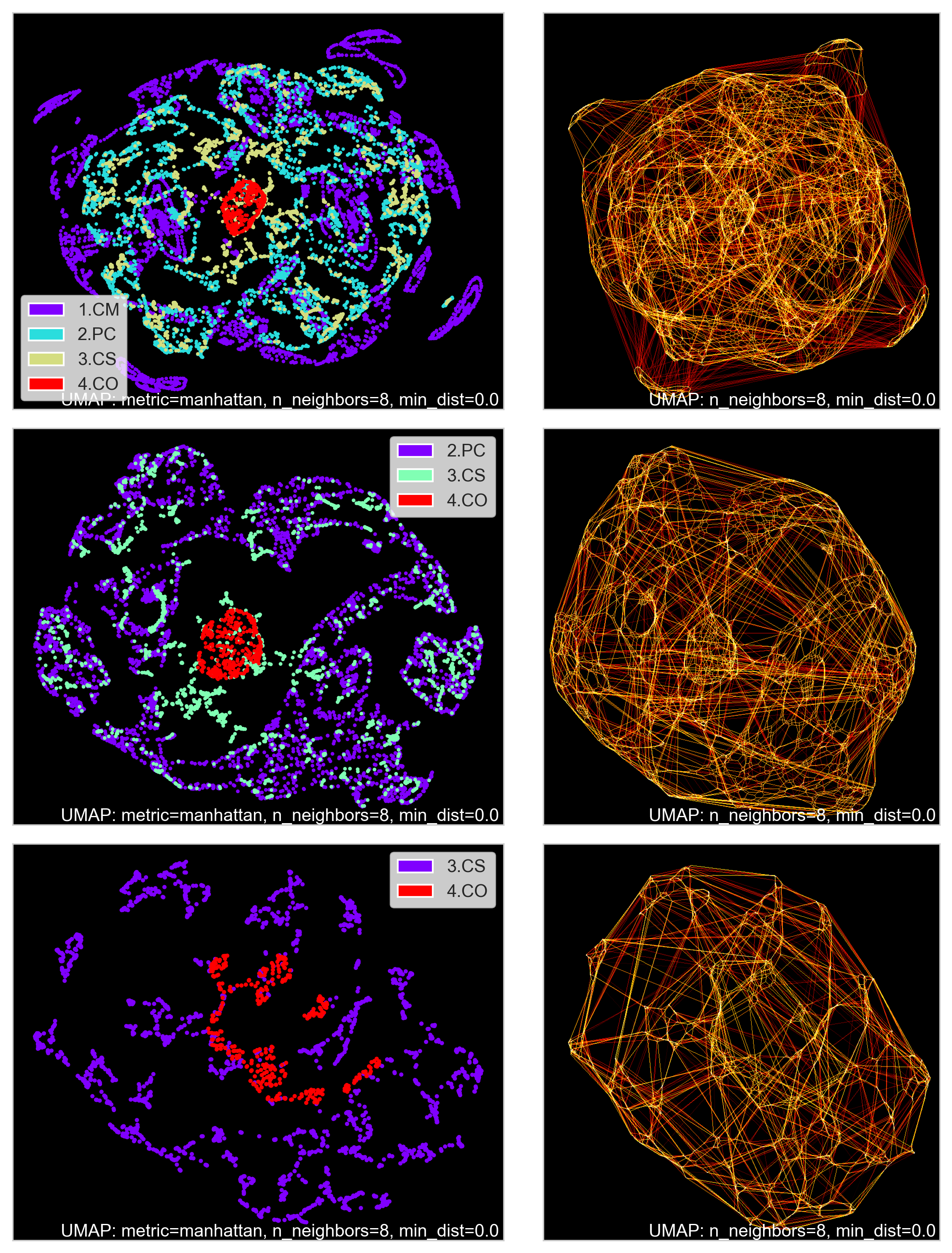}
\end{center}

\subsection{15\% Cytoskeleton Volume}
\begin{center} 
\includegraphics[width=0.8\textwidth]{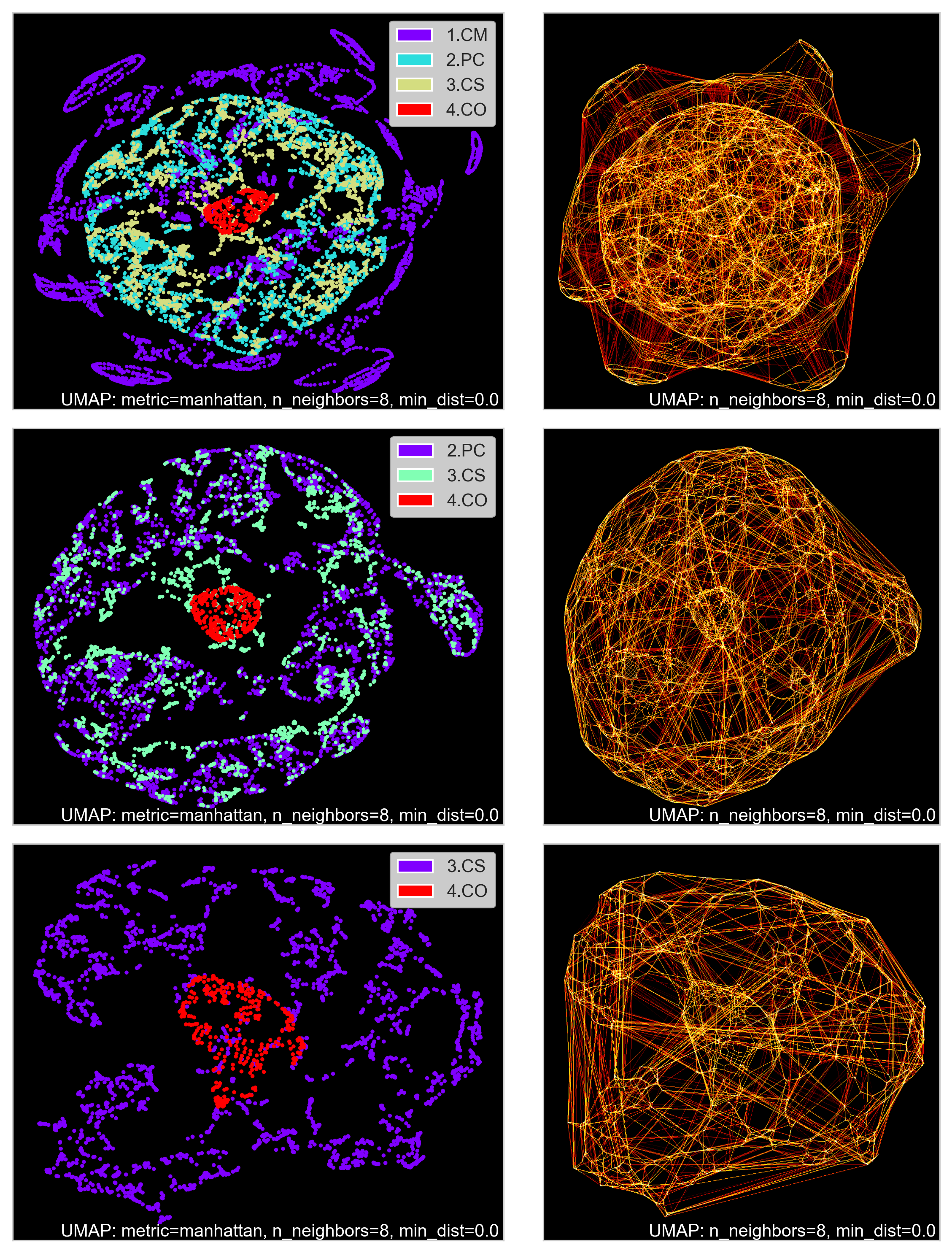}
\end{center}

\newpage
\section{Signal Map of 3D Cell-Reservoir for Various Cytoskeleton Volume and Source Distribution\label{sec:sm2}}
\subsection{Point Source}
\begin{center}
\includegraphics[width=0.7\textwidth]{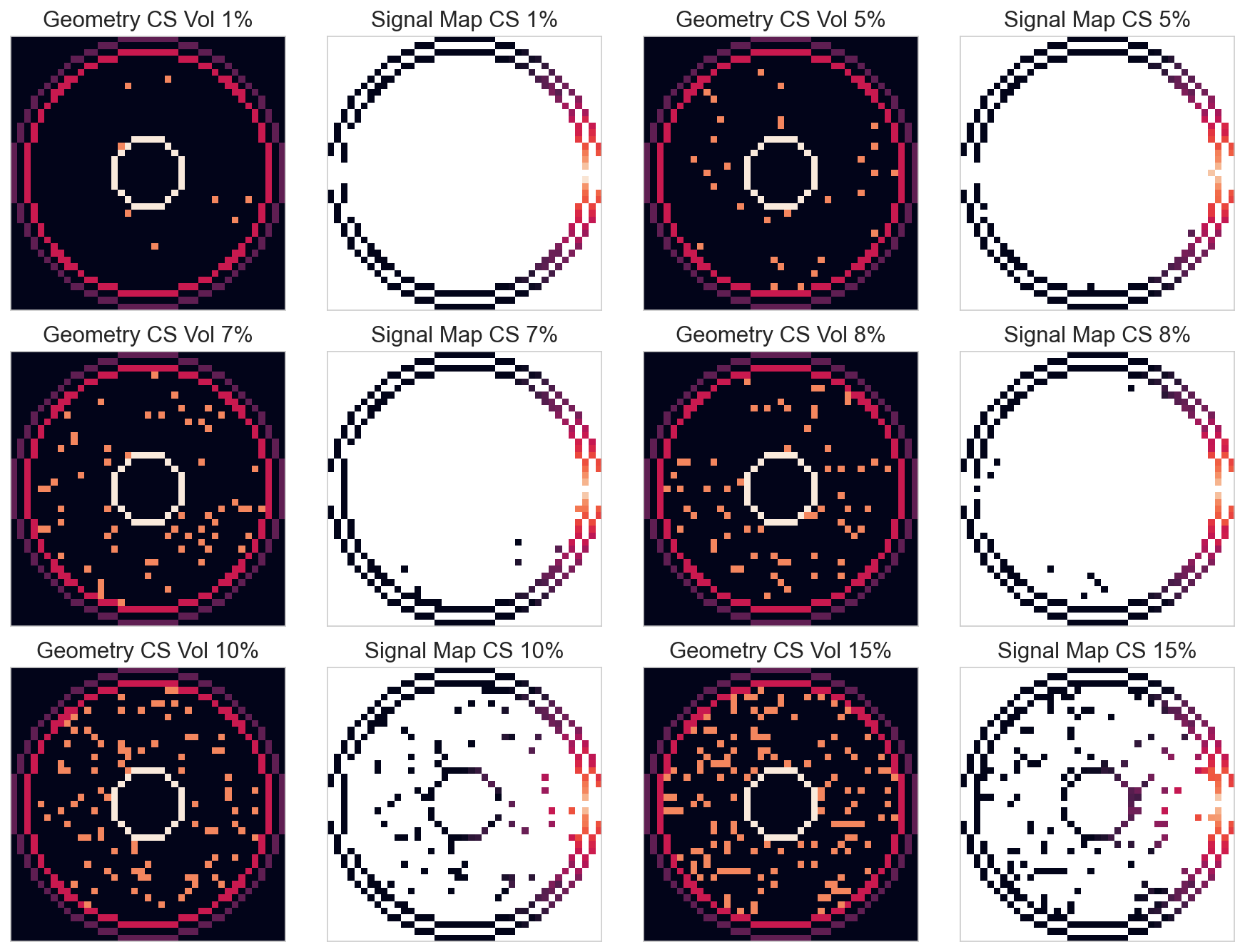}
\end{center}

\subsection{Spherical Source}
\begin{center}
\includegraphics[width=0.7\textwidth]{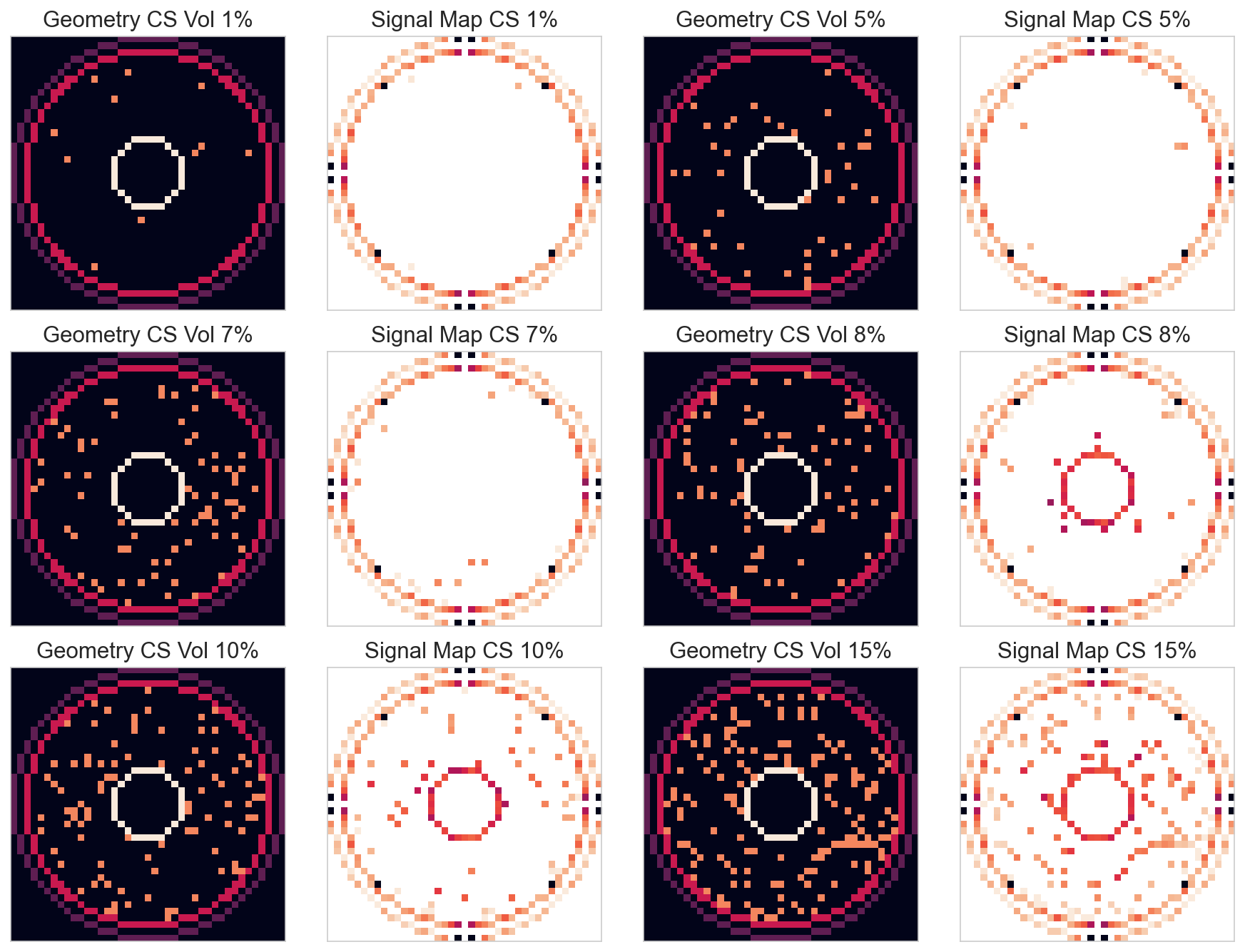}
\end{center}

\newpage
\section{Cell-Reservoir Response for Various Noise Level and Source Distribution\label{sec:sm3}}
\subsection{Point Source, 0.05 Noise Level}
\begin{center}
\includegraphics[width=0.6\textwidth]{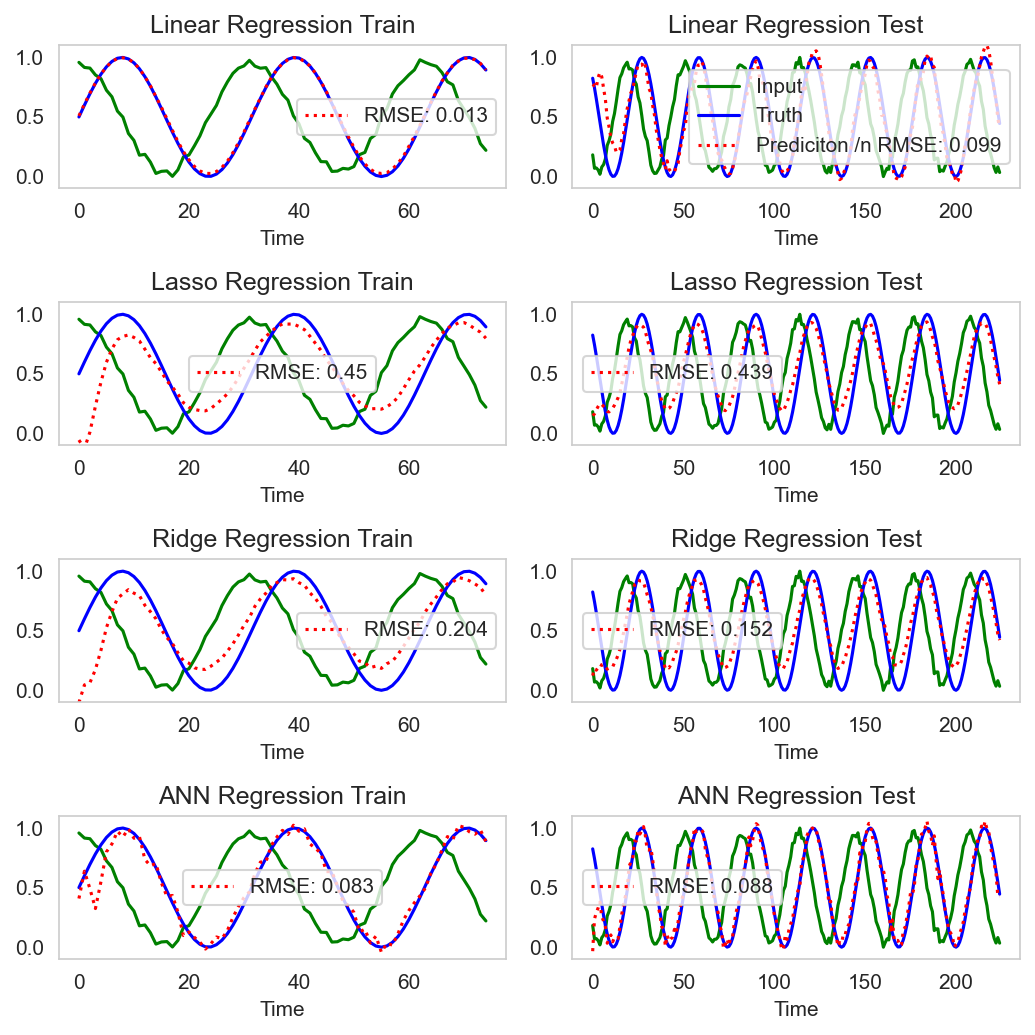} 
\includegraphics[width=0.6\textwidth]{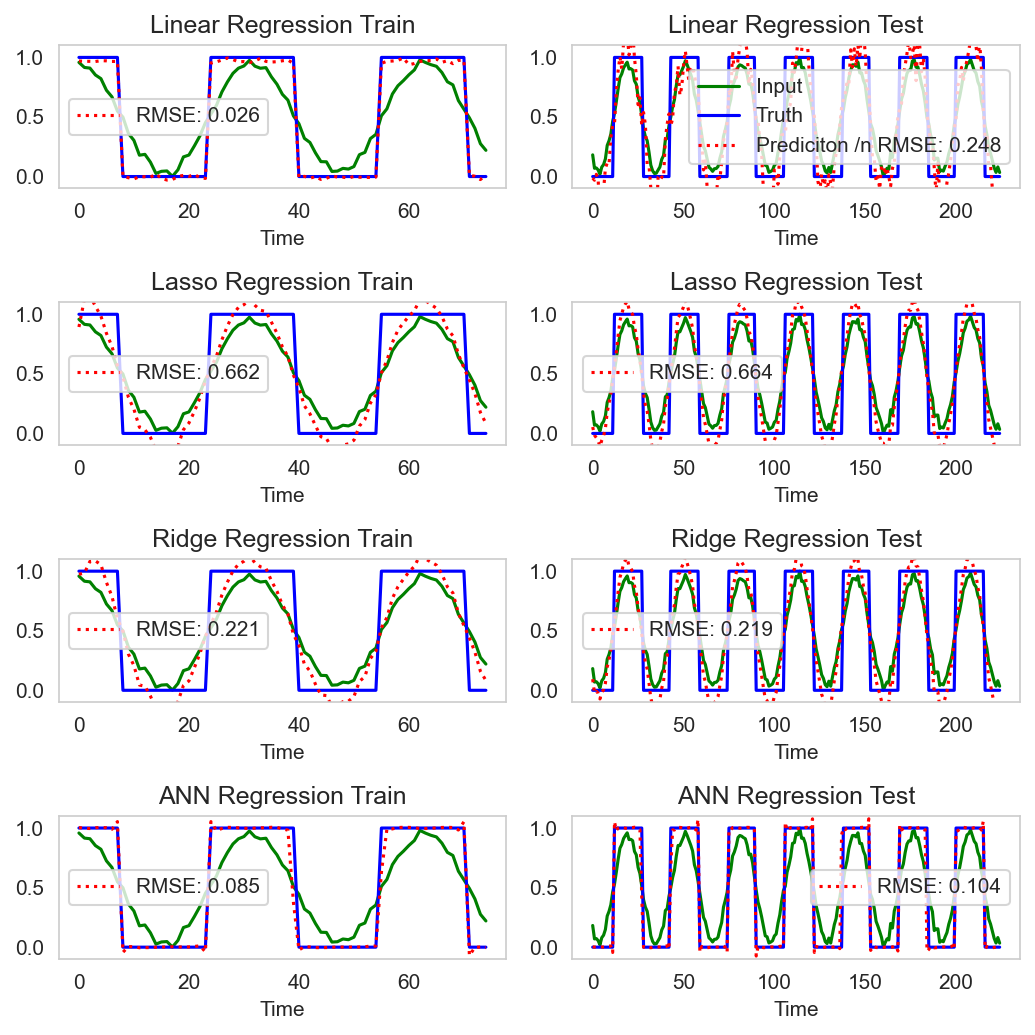}
\end{center}

\subsection{Point Source, 0.1 Noise Level}
\begin{center}
\includegraphics[width=0.6\textwidth]{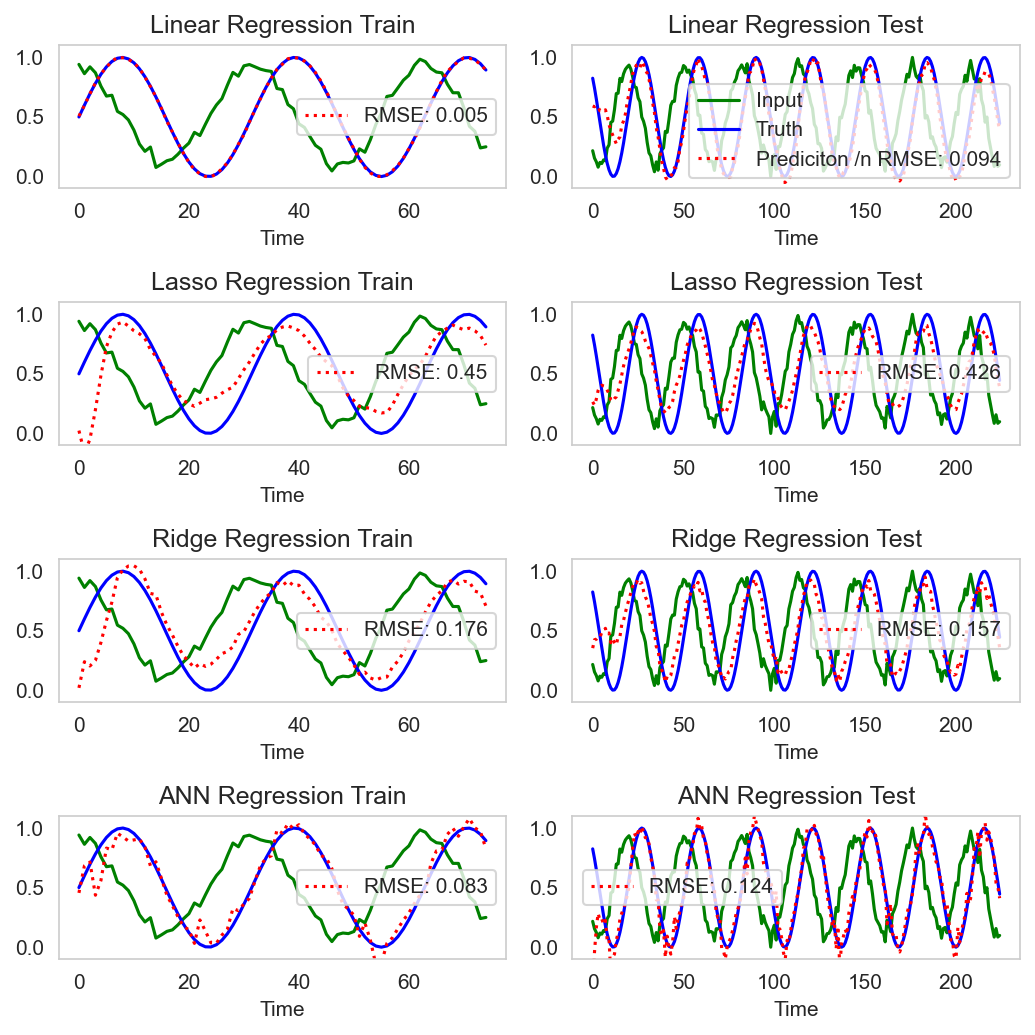}
\includegraphics[width=0.6\textwidth]{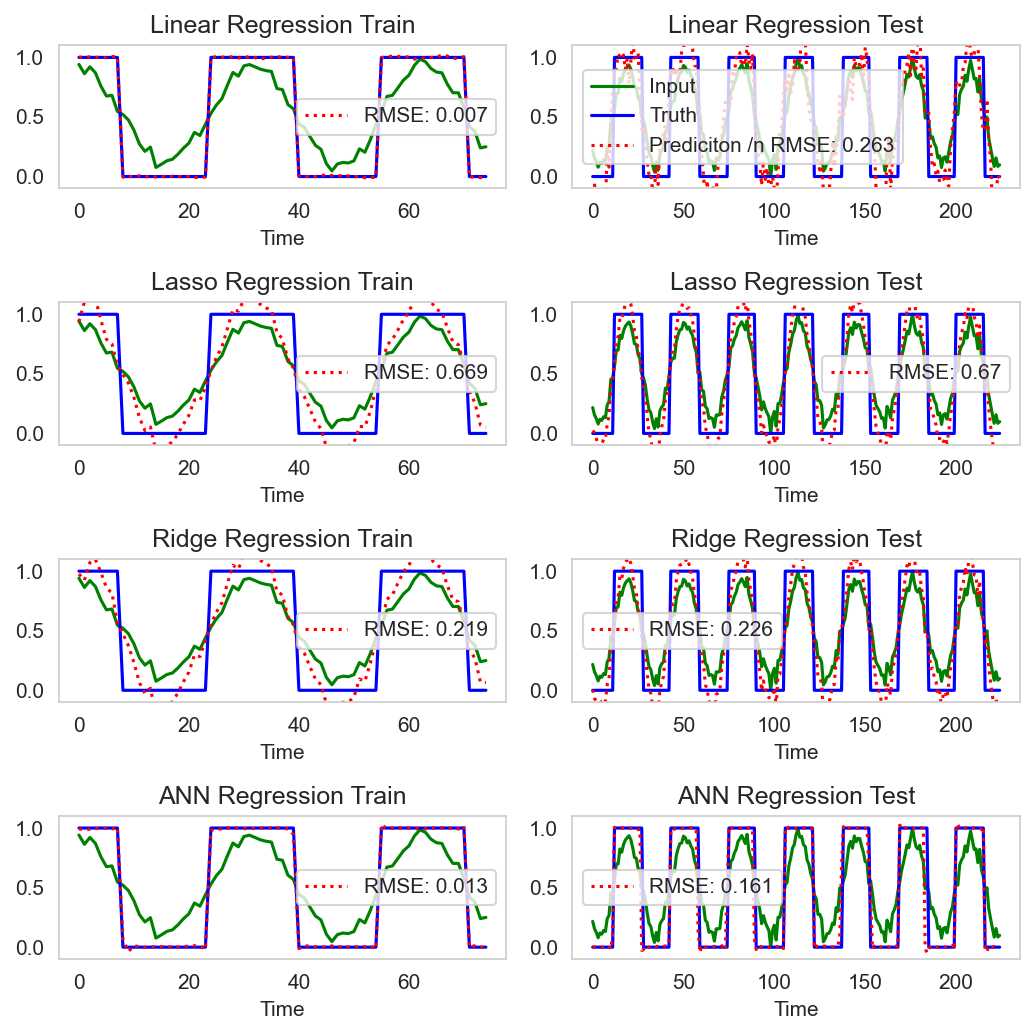}
\end{center}

\subsection{Point Source, 0.25 Noise Level}
\begin{center}
\includegraphics[width=0.6\textwidth]{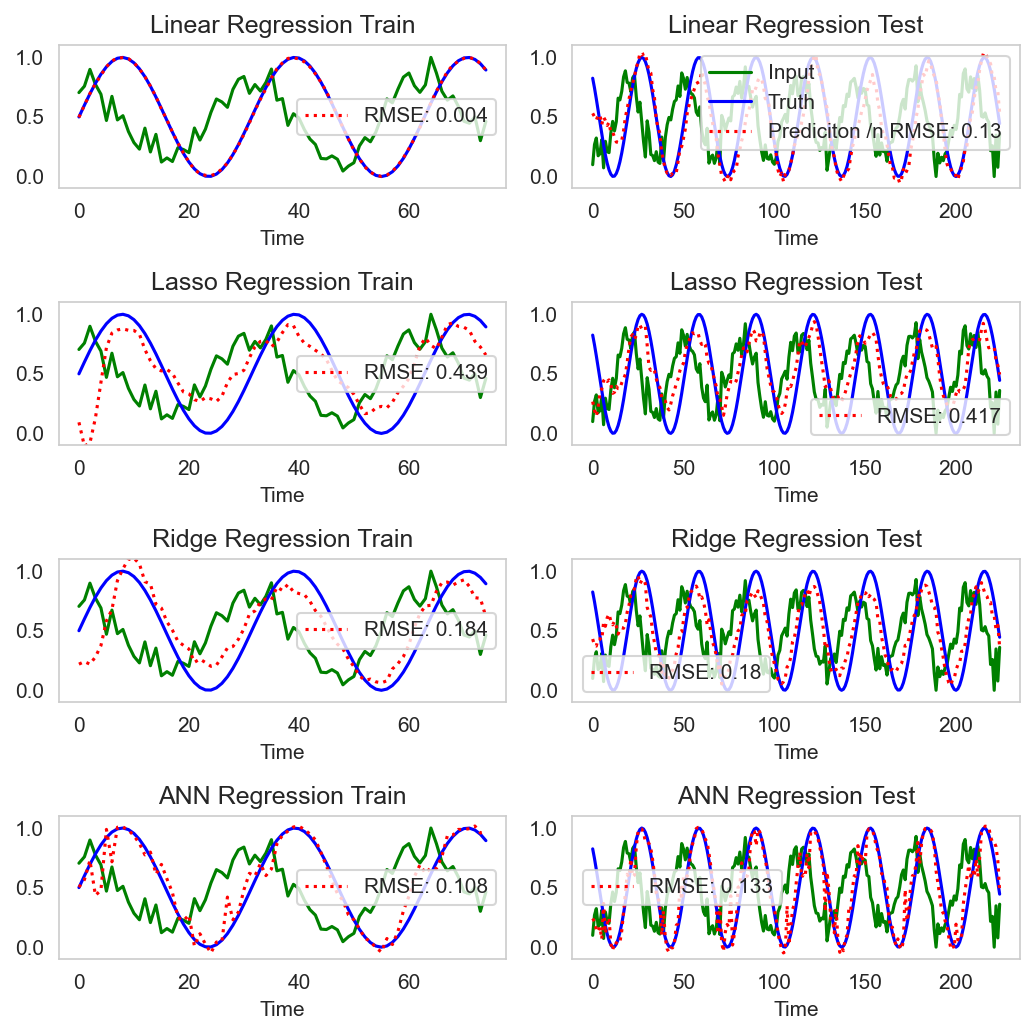}
\includegraphics[width=0.6\textwidth]{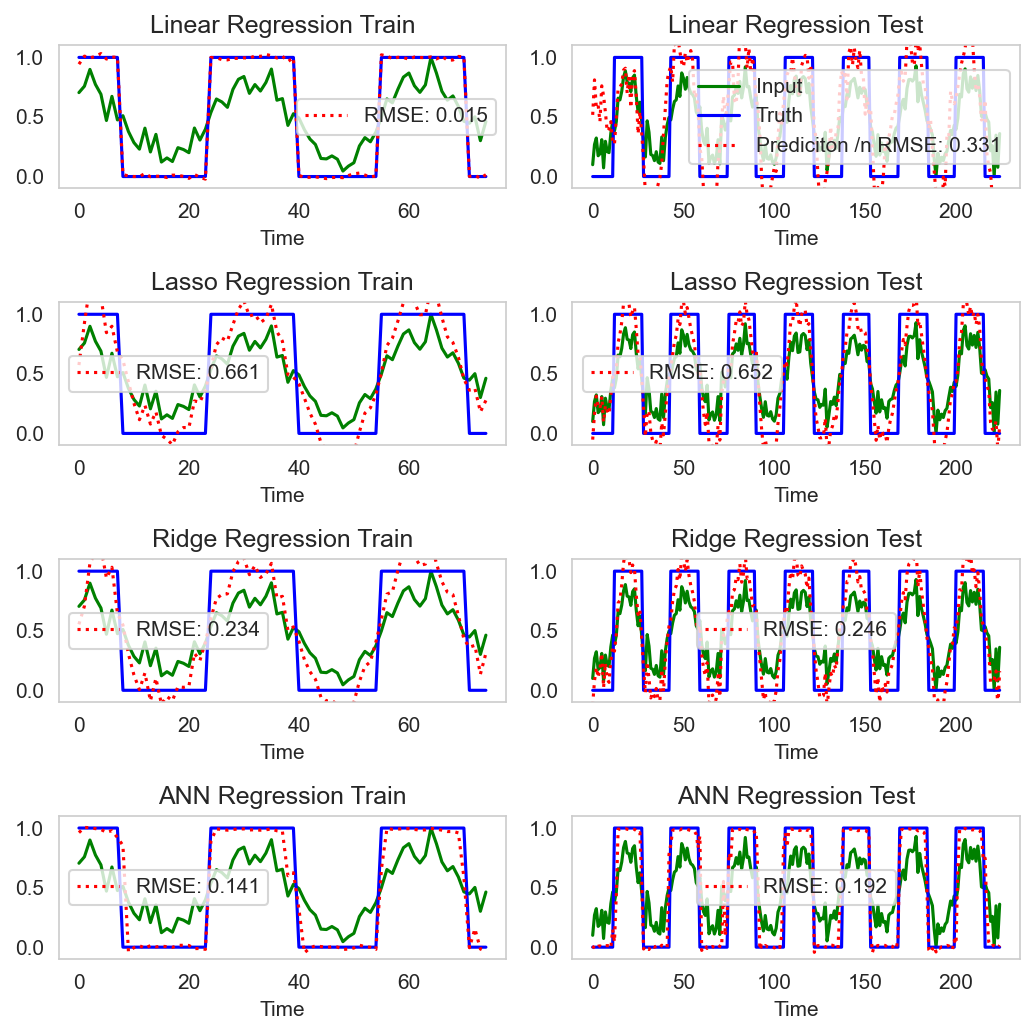}
\end{center}

\subsection{Point Source, 0.5 Noise Level}
\begin{center}
\includegraphics[width=0.6\textwidth]{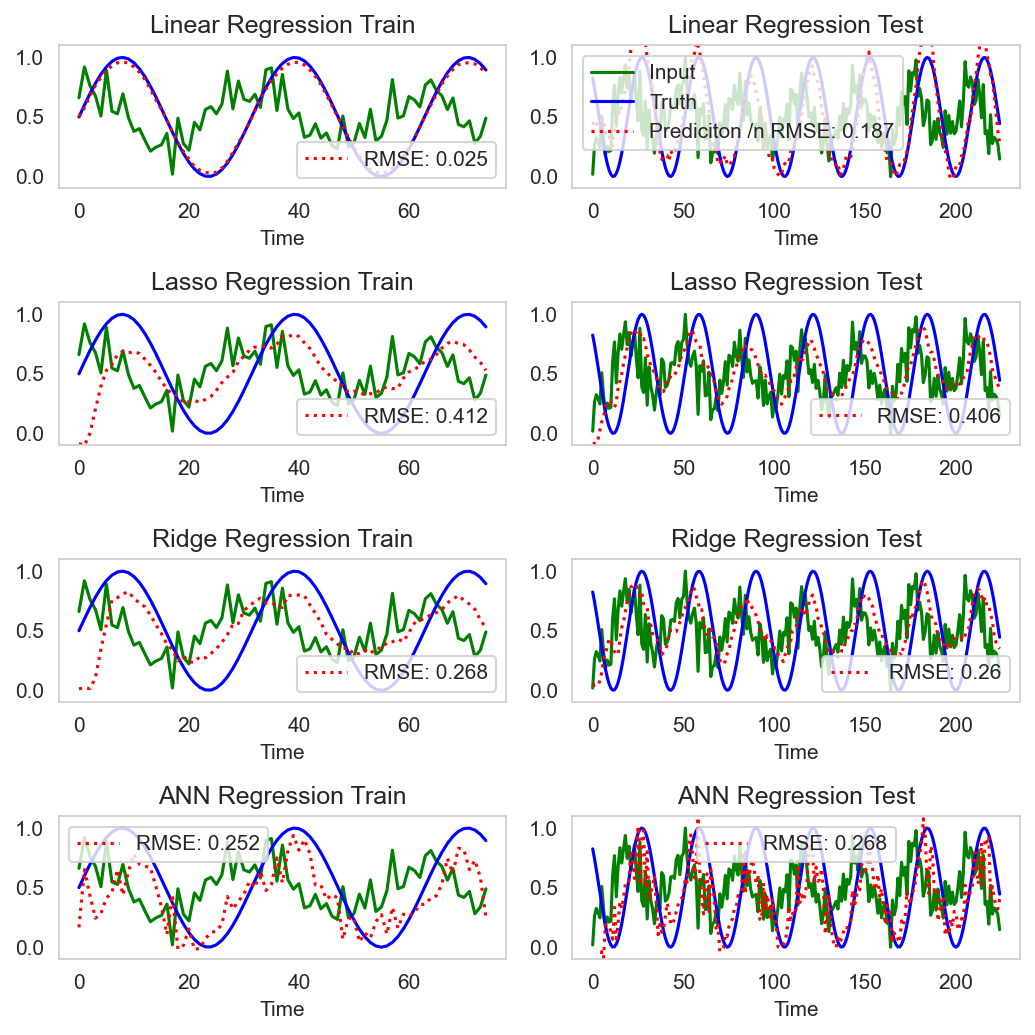}
\includegraphics[width=0.6\textwidth]{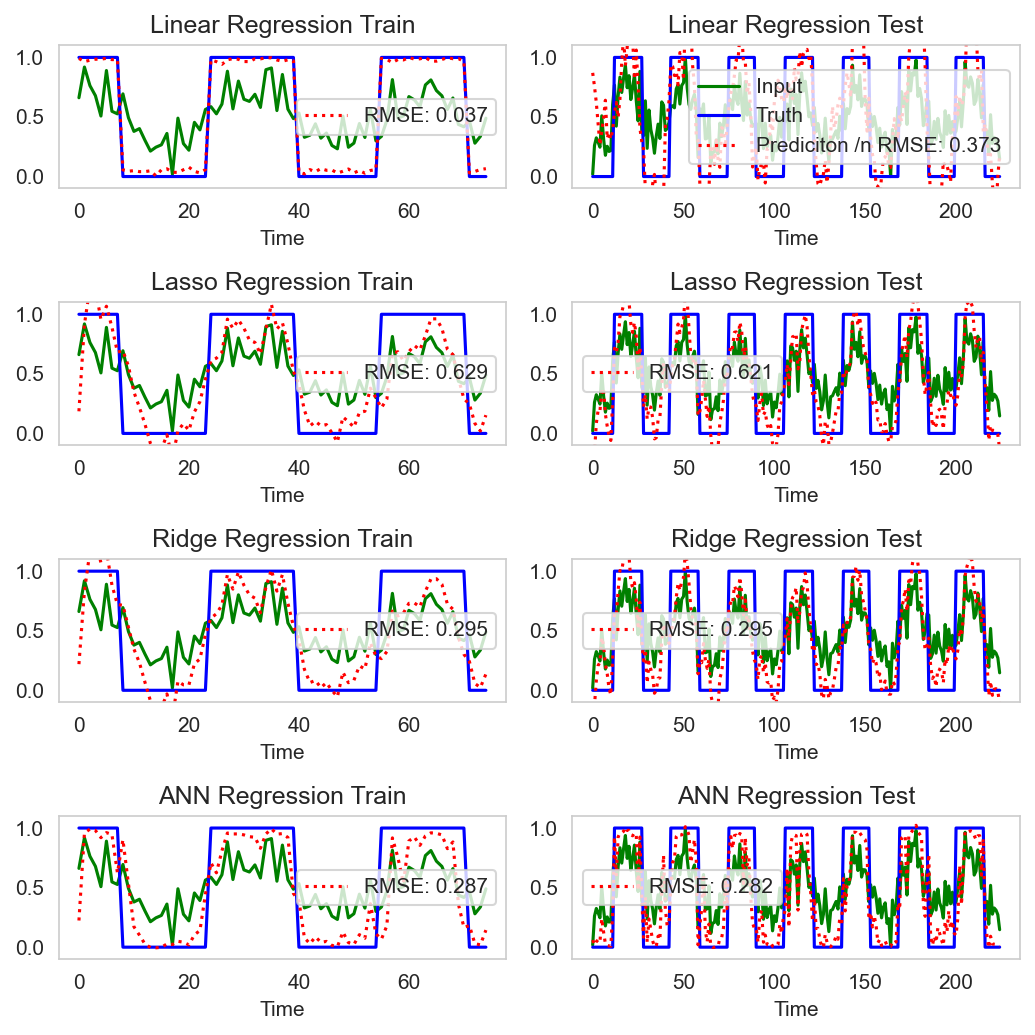}
\end{center}

\subsection{Point Source, 1.0 Noise Level}
\begin{center}
\includegraphics[width=0.6\textwidth]{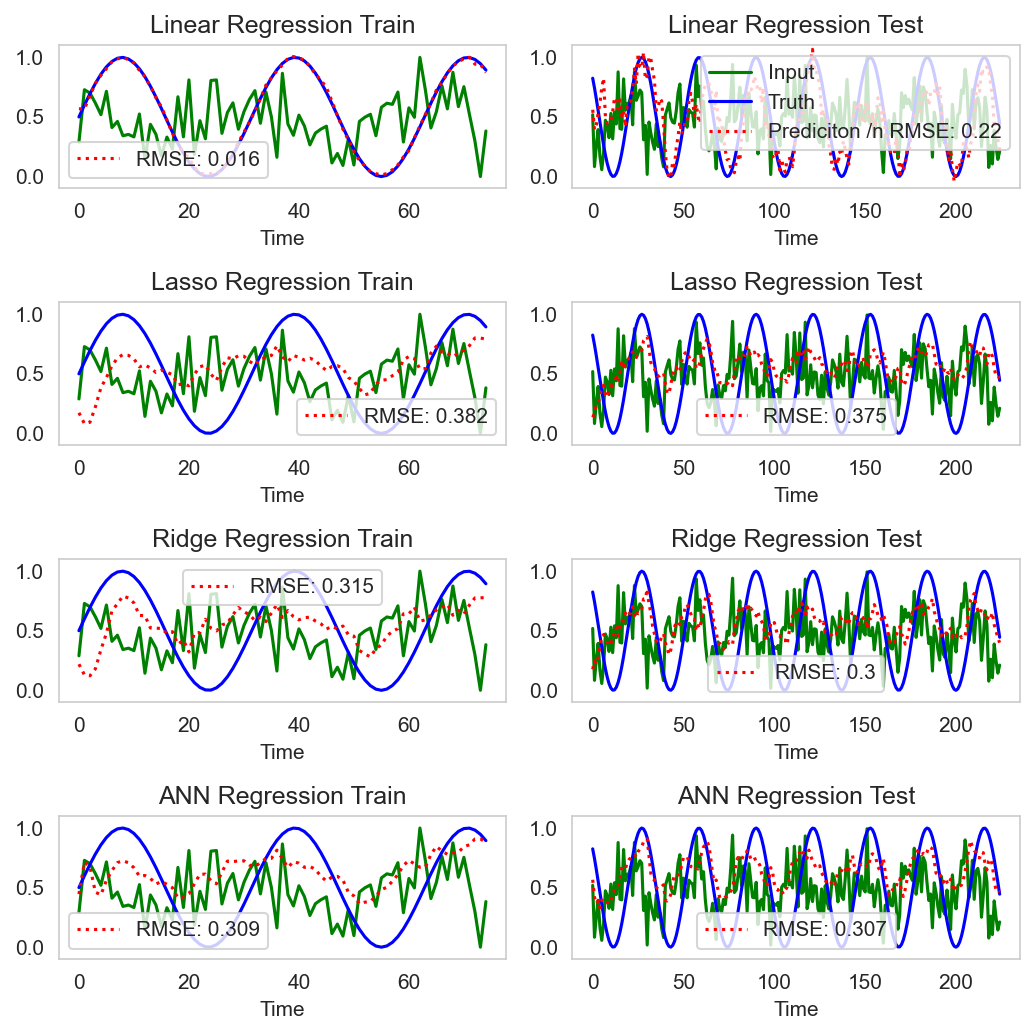}
\includegraphics[width=0.6\textwidth]{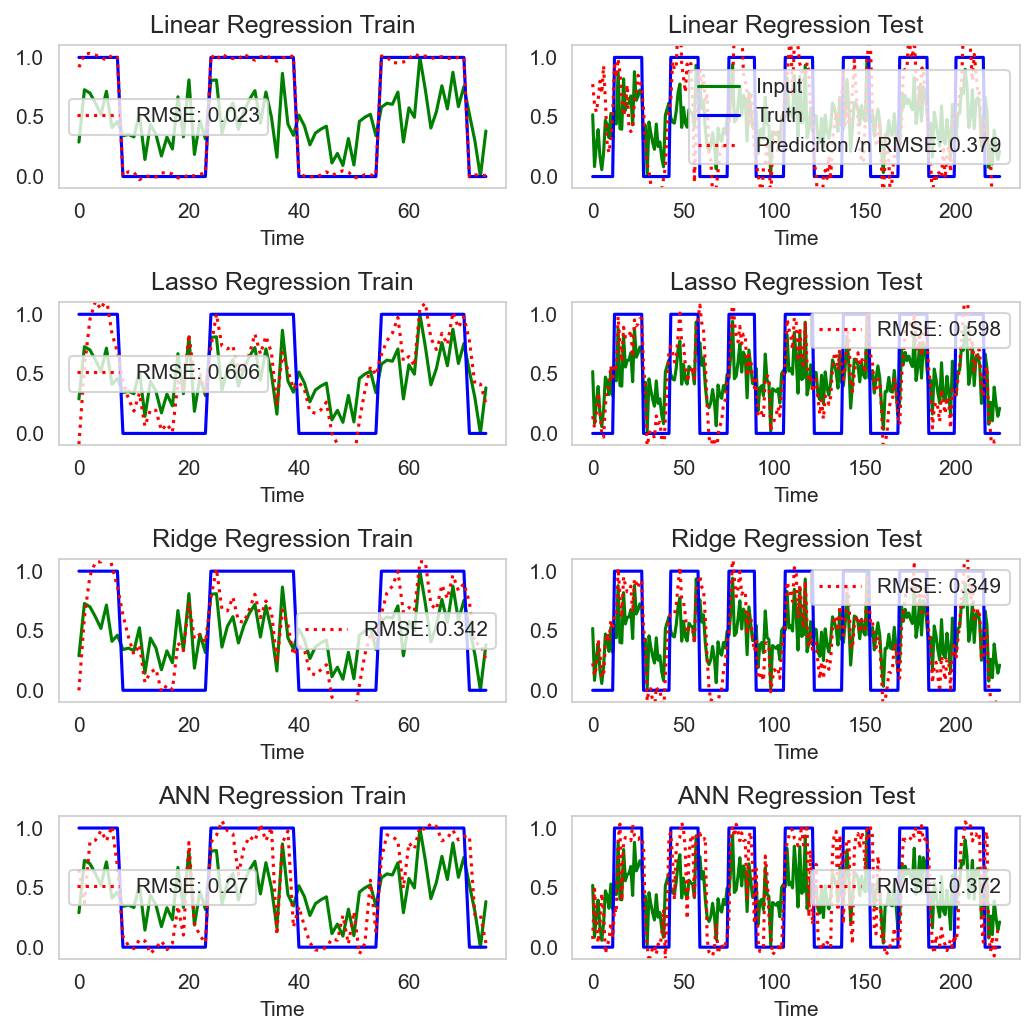}
\end{center}

\subsection{Spherical Source, 0.05 Noise Level}
\begin{center}
\includegraphics[width=0.5\textwidth]{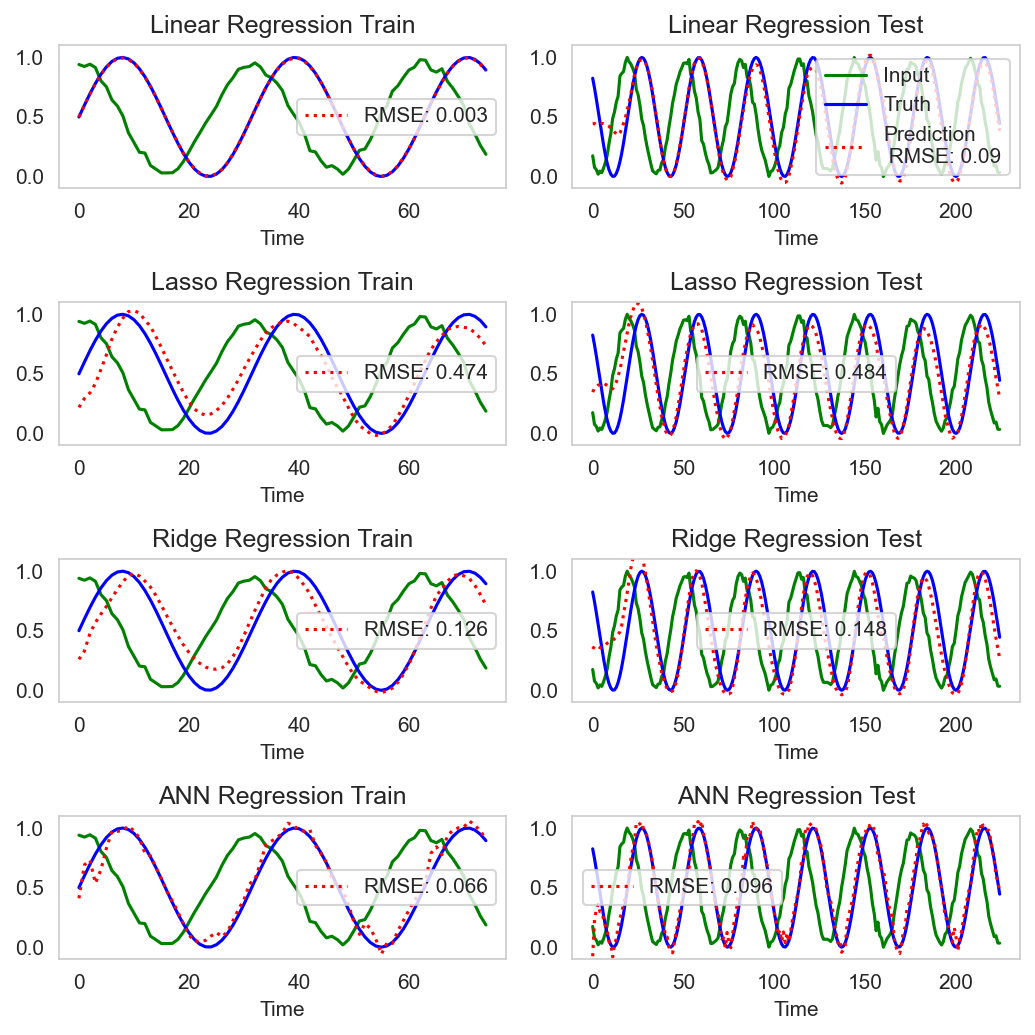}
\includegraphics[width=0.5\textwidth]{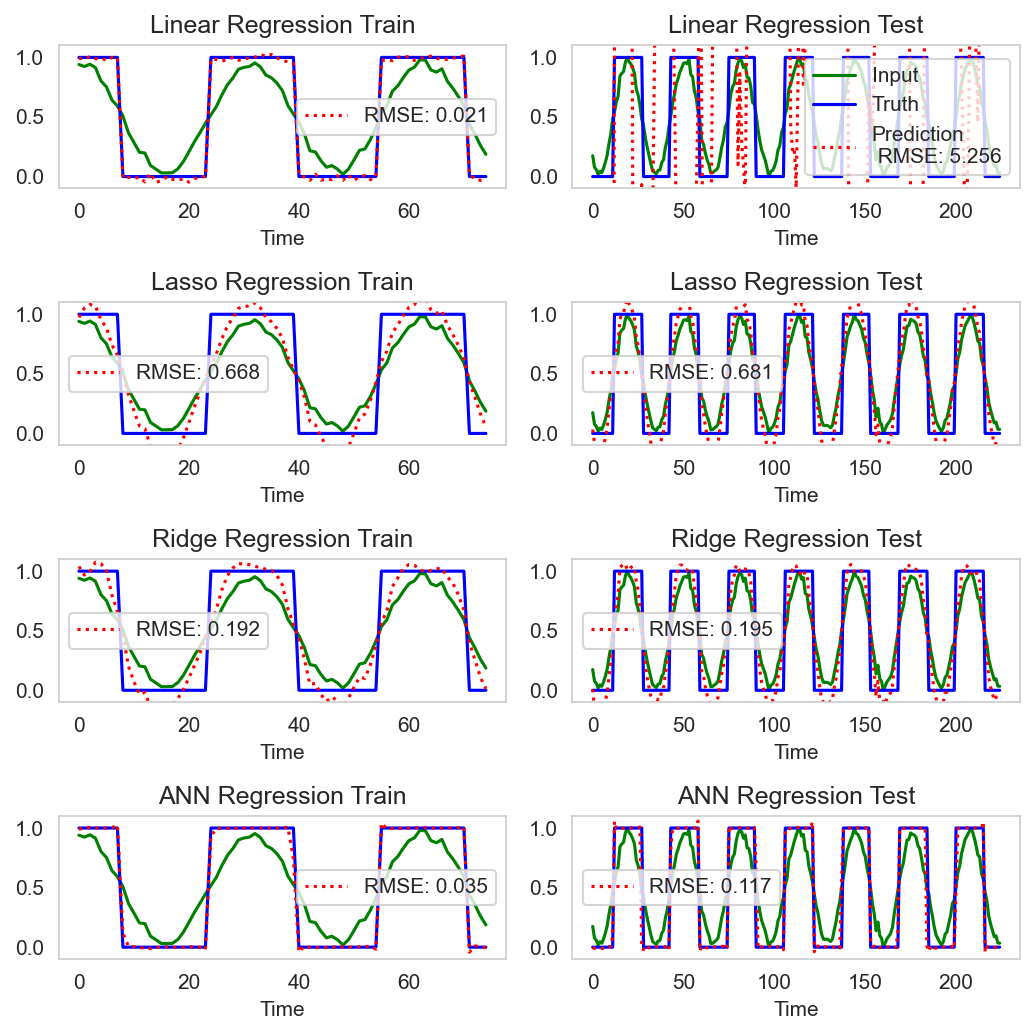}
\end{center}

\subsection{Spherical Source, 0.1 Noise Level}
\begin{center}
\includegraphics[width=0.6\textwidth]{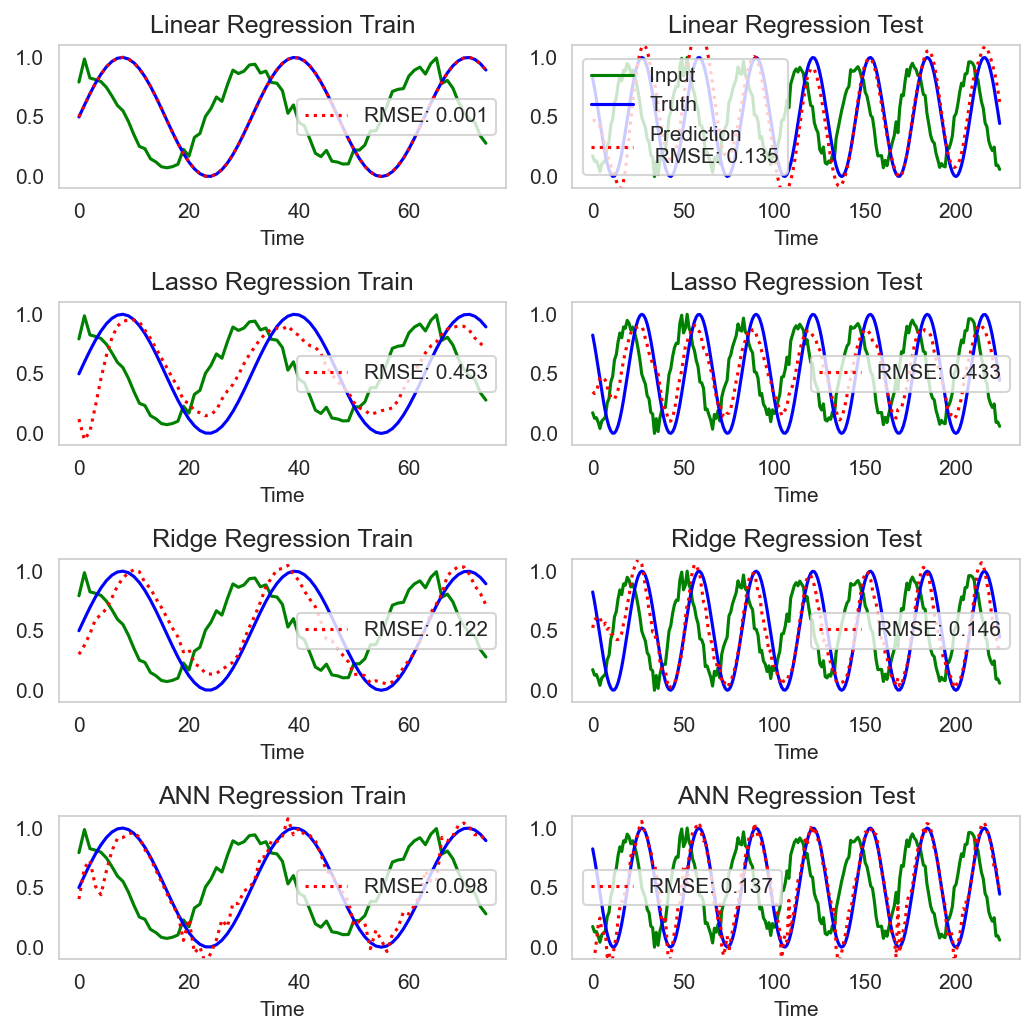}
\includegraphics[width=0.6\textwidth]{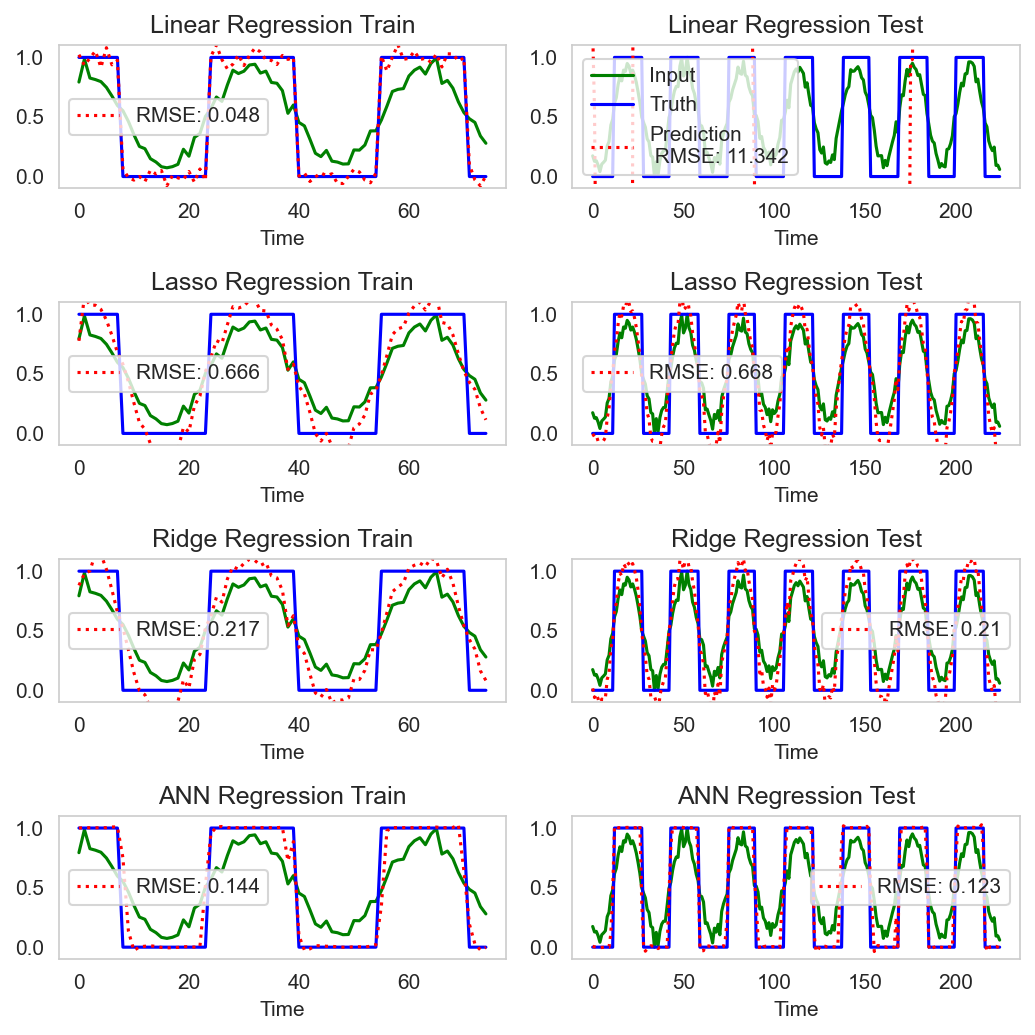}
\end{center}

\subsection{Spherical Source, 0.25 Noise Level}
\begin{center}
\includegraphics[width=0.6\textwidth]{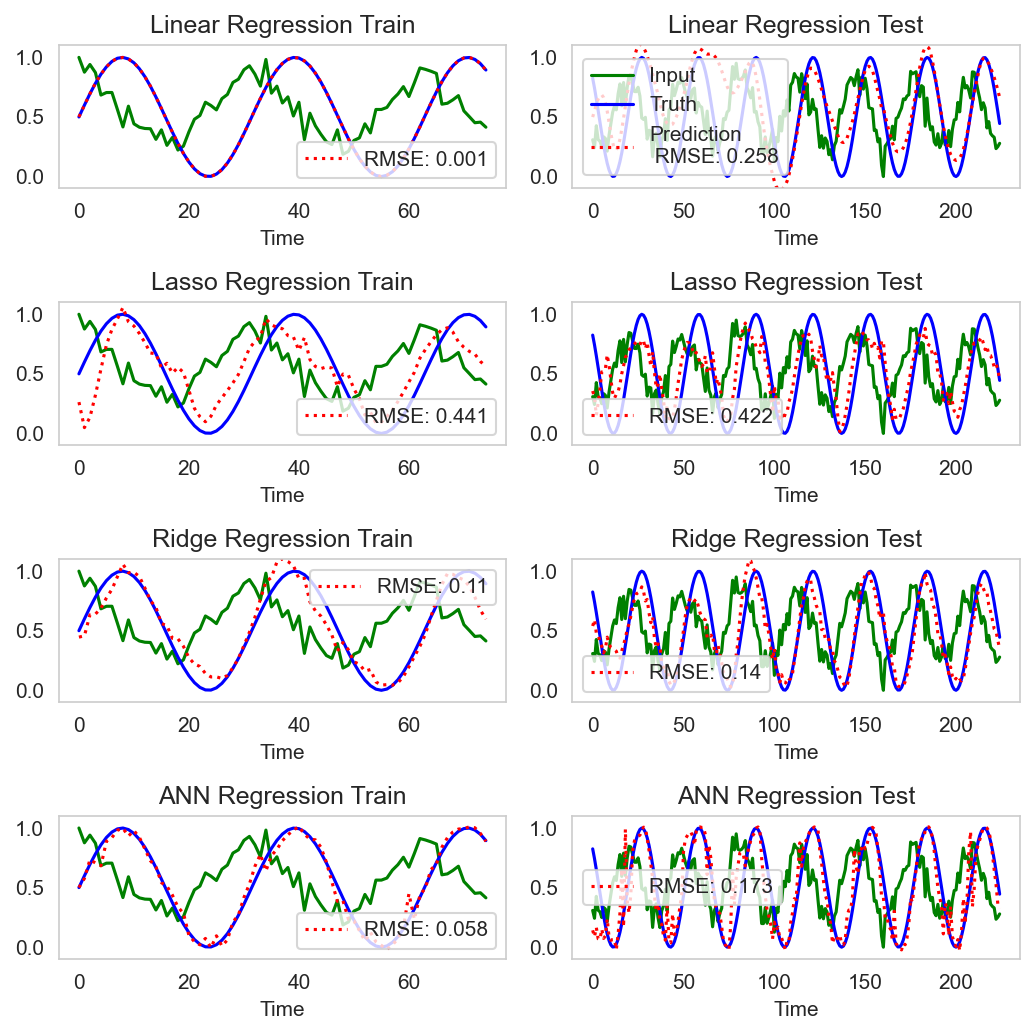}
\includegraphics[width=0.6\textwidth]{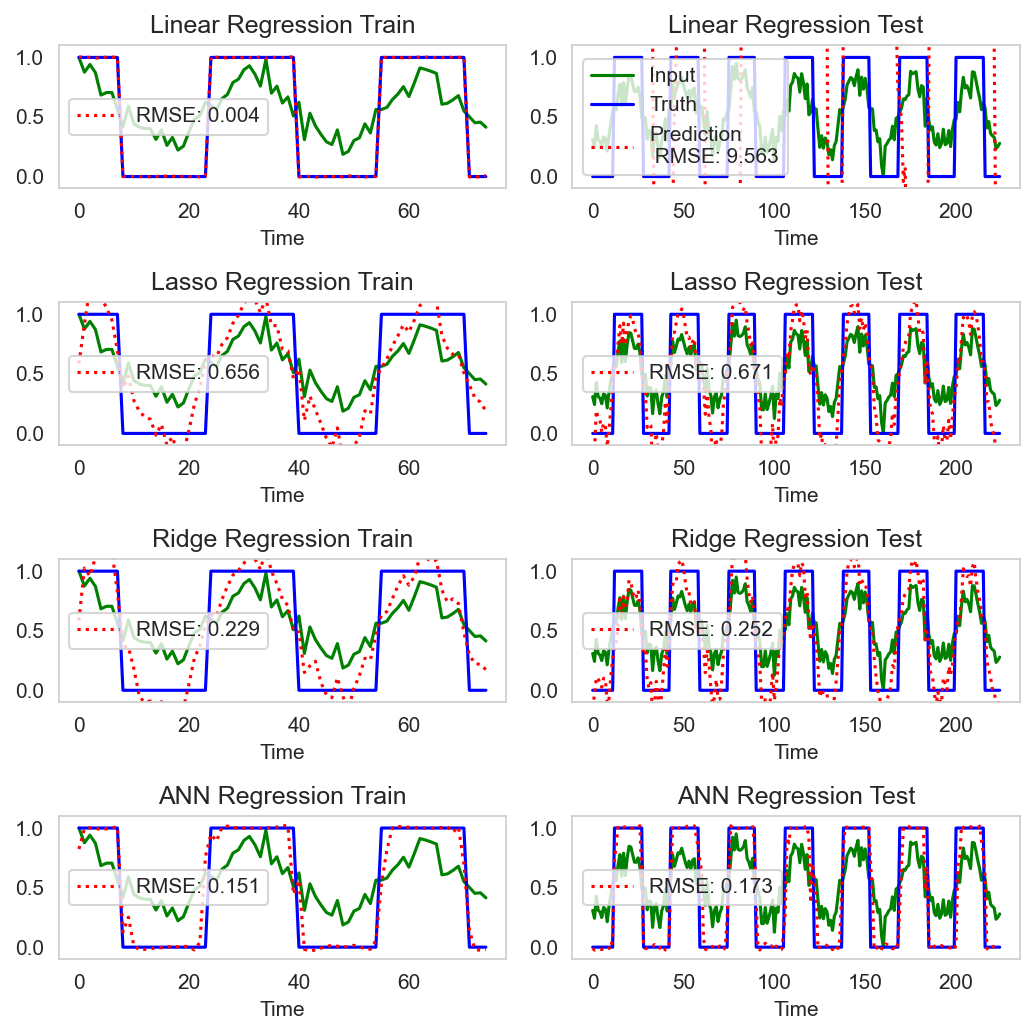}
\end{center}

\subsection{Spherical Source, 0.5 Noise Level}
\begin{center}
\includegraphics[width=0.6\textwidth]{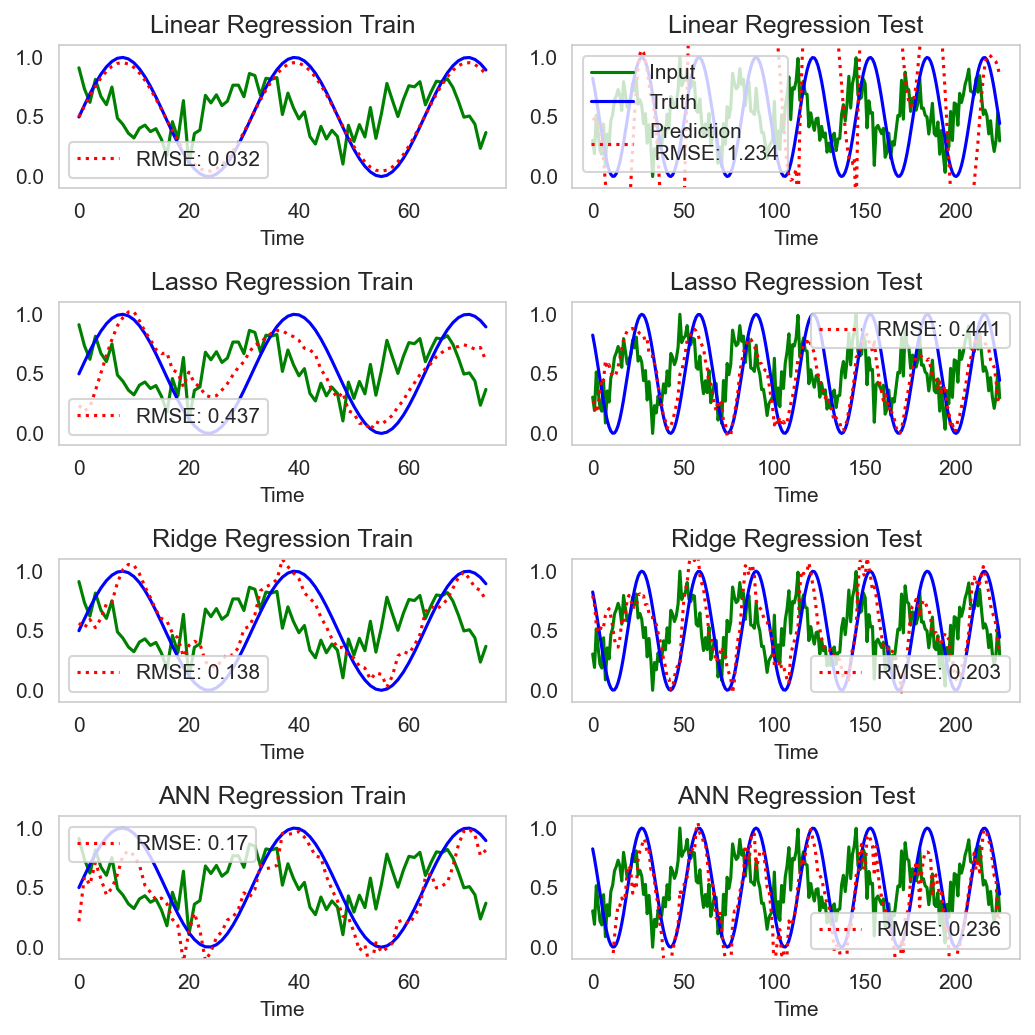}
\includegraphics[width=0.6\textwidth]{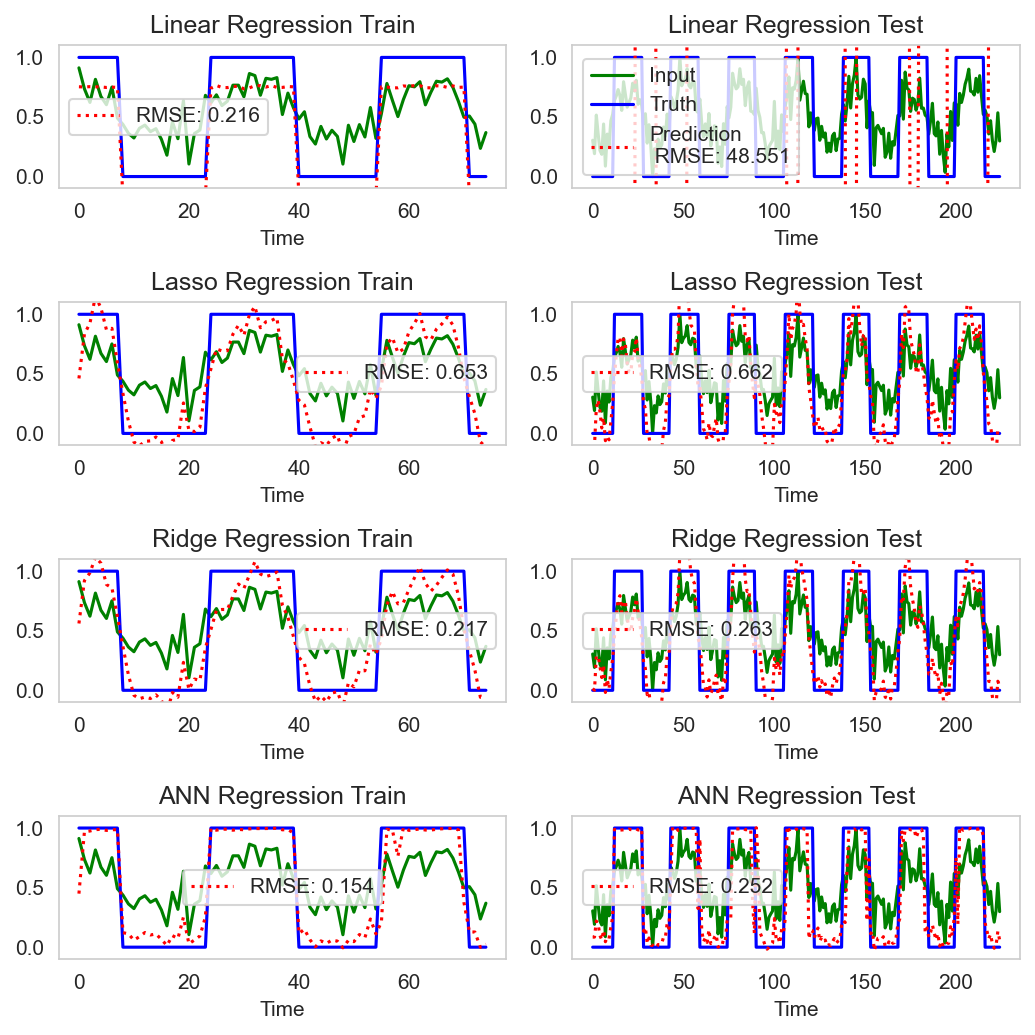}
\end{center}

\subsection{Spherical  Source, 1.0 Noise Level}
\begin{center}
\includegraphics[width=0.6\textwidth]{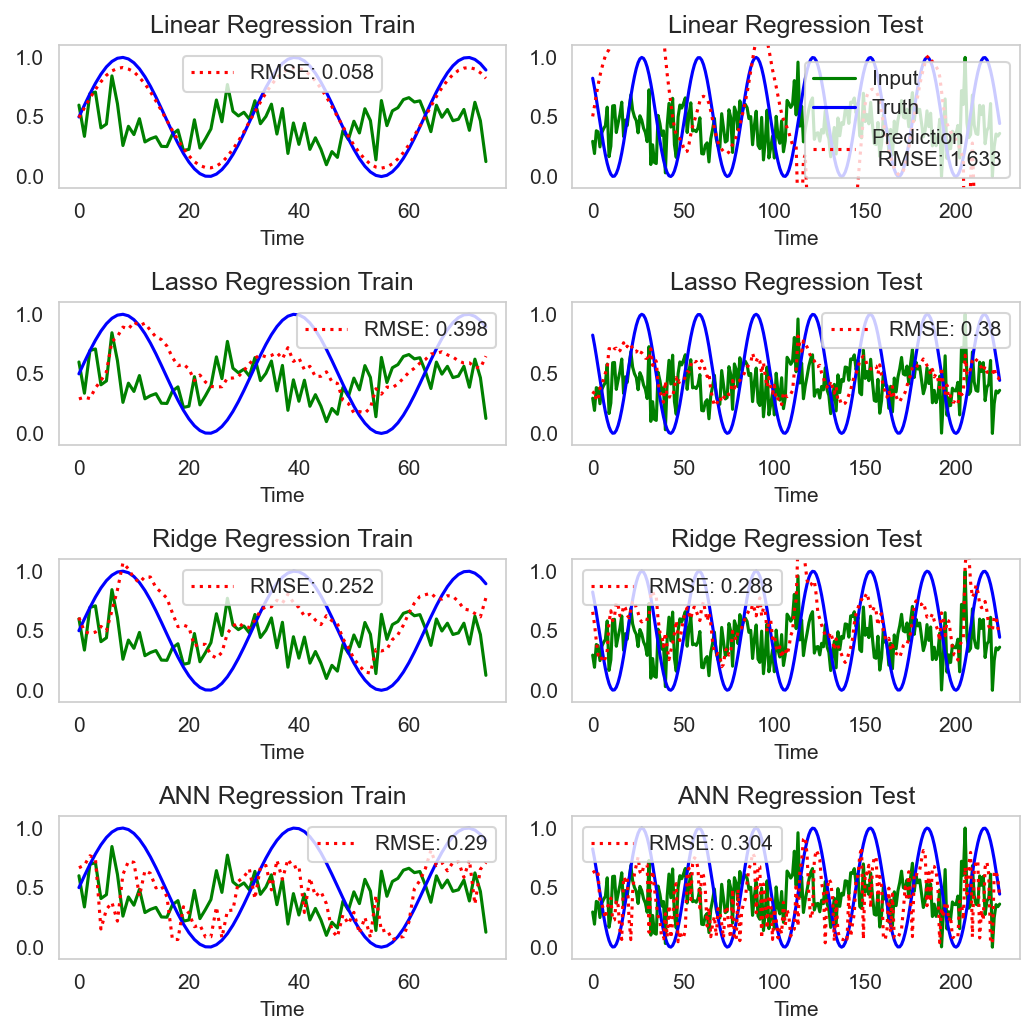}
\includegraphics[width=0.6\textwidth]{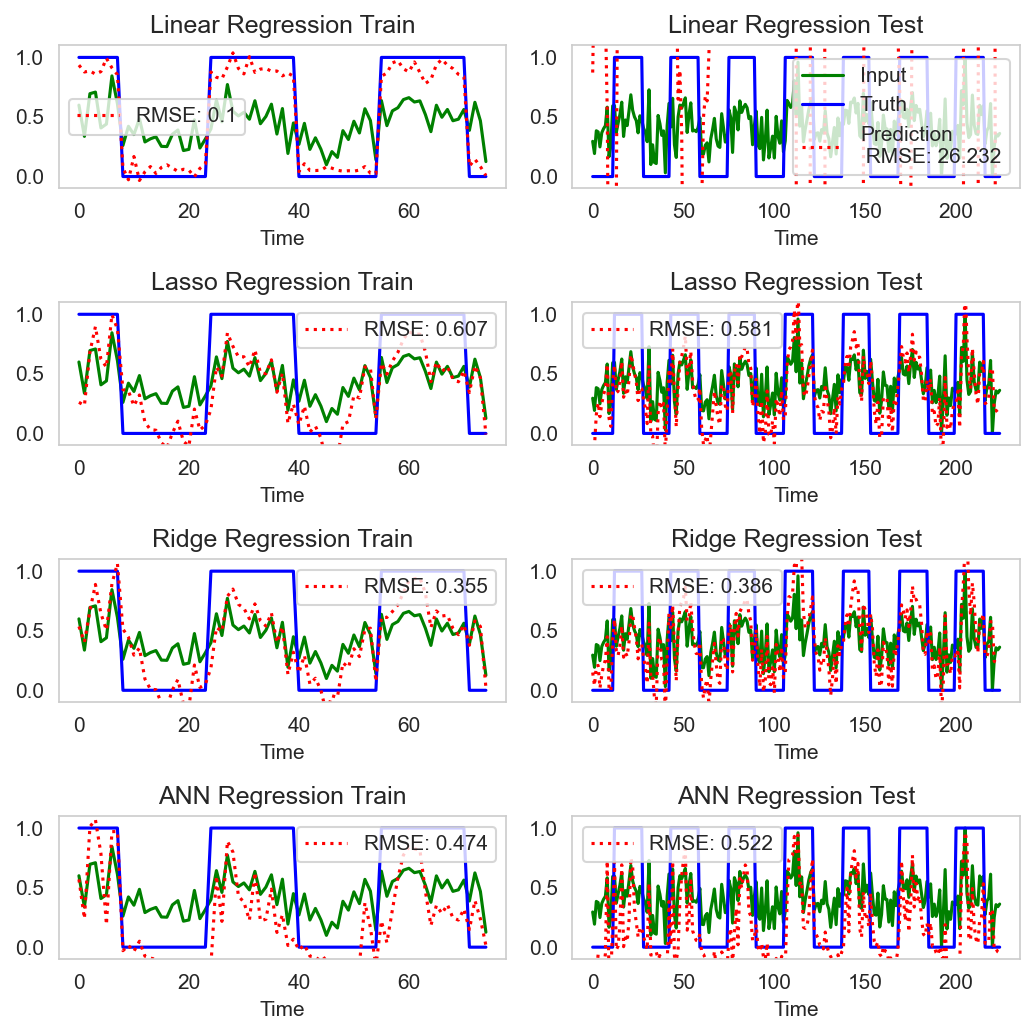}
\end{center}

\end{document}